%% file: pbh_pe_prl.tex
\documentclass[
 amsmath,amssymb,
 prl,aps,superscriptaddress,twocolumn,
floatfix,
]{revtex4-2}

\usepackage{aas_macros}
\usepackage{graphicx}
\usepackage{amsmath}
\usepackage[caption=false]{subfig}
\usepackage{siunitx}
\usepackage{placeins}
\usepackage{color}
\usepackage{standalone}
\usepackage{dcolumn}
\usepackage{tensor}
\usepackage{bm}
\usepackage{microtype}
\usepackage{etoolbox}
\usepackage{amssymb}
\usepackage{mathrsfs}
\usepackage{accents}
\usepackage[normalem]{ulem}
\usepackage{soul} 
\usepackage[dvipsnames]{xcolor}
\usepackage[colorlinks,urlcolor=NavyBlue,citecolor=NavyBlue,linkcolor=NavyBlue,pdfusetitle]{hyperref}
\usepackage[all]{hypcap}
\usepackage[inline]{enumitem}
\usepackage[utf8]{inputenc}
\usepackage{xspace}
\usepackage[printonlyused, nolist]{acronym}
\usepackage{lineno}
\usepackage{hyperref}

\newcommand{\Mc}{\ensuremath{\mathcal{M}_c}\xspace}
\newcommand{\Mtotal}{\ensuremath{M_{\rm tot}}\xspace}
\newcommand{\msun}{\ensuremath{M_{\odot}}\xspace}

\newcommand{\zcritvalue}{\ensuremath{30}\xspace}
\newcommand{\zcrit}{\ensuremath{z_\mathrm{crit}}\xspace}
\newcommand{\fcrit}{\ensuremath{f^{\mathrm{PBH}}_{\mathrm{III}}}\xspace}
\newcommand{\ztrue}{\ensuremath{z_{\mathrm{true}}}\xspace}

\newcommand{\aIII}{\ensuremath{a_{\rm III}}\xspace}
\newcommand{\bIII}{\ensuremath{b_{\rm III}}\xspace}
\newcommand{\zIII}{\ensuremath{z_{\rm III}}\xspace}
\newcommand{\npbh}{\ensuremath{\dot{n}_{\rm PBH}}\xspace}
\newcommand{\nIII}{\ensuremath{\dot{n}_{\rm III}}\xspace}
\newcommand{\ptot}{\ensuremath{p_{\rm tot}}\xspace}
\newcommand{\BF}{\ensuremath{\mathcal{B}^{\mathrm{P}}_0\xspace}}
\newcommand{\supmat}{Supplementary Material\xspace}
\newcommand{\letter}{\textit{Letter}\xspace}
\newcommand{\prlsection}[1]{\textit{#1}  -- \xspace}
\newcommand{\penncosmos}{\affiliation{Institute for Gravitation and the Cosmos, Department of Physics, Pennsylvania State University, University Park, PA, 16802, USA}}

\newcommand{\pennastro}{\affiliation{Department of Astronomy \& Astrophysics, Pennsylvania State University, University Park, PA, 16802, USA}}
\newcommand{\cardiff}{\affiliation{School of Physics and Astronomy, Cardiff University, Cardiff, UK, CF24 3AA}
}

\begin{document}
\title{On the single-event-based identification of primordial black hole mergers at cosmological distances}

\author{Ken K.~Y.~Ng}
\email{kenkyng@mit.edu}
\affiliation{LIGO, Massachusetts Institute of Technology, Cambridge, Massachusetts 02139, USA}
\affiliation{Kavli Institute for Astrophysics and Space Research, Massachusetts Institute of Technology, Cambridge, Massachusetts 02139, USA}
\affiliation{Department of Physics, Massachusetts Institute of Technology, Cambridge, Massachusetts 02139, USA}
\author{Shiqi Chen}
\affiliation{LIGO, Massachusetts Institute of Technology, Cambridge, Massachusetts 02139, USA}
\affiliation{Kavli Institute for Astrophysics and Space Research, Massachusetts Institute of Technology, Cambridge, Massachusetts 02139, USA}
\affiliation{Department of Physics, Massachusetts Institute of Technology, Cambridge, Massachusetts 02139, USA}
\author{Boris Goncharov}
\affiliation{Gran Sasso Science Institute (GSSI), I-67100 L'Aquila, Italy}
\affiliation{INFN, Laboratori Nazionali del Gran Sasso, I-67100 Assergi, Italy}
\author{Ulyana Dupletsa}
\affiliation{Gran Sasso Science Institute (GSSI), I-67100 L'Aquila, Italy}
\affiliation{INFN, Laboratori Nazionali del Gran Sasso, I-67100 Assergi, Italy}
\author{Ssohrab Borhanian}
\penncosmos
\affiliation{Theoretisch-Physikalisches Institut, Friedrich-Schiller-Universit\"at Jena, 07743, Jena, Germany}

\author{Marica Branchesi}
\affiliation{Gran Sasso Science Institute (GSSI), I-67100 L'Aquila, Italy}
\affiliation{INFN, Laboratori Nazionali del Gran Sasso, I-67100 Assergi, Italy}
\author{Jan Harms}
\affiliation{Gran Sasso Science Institute (GSSI), I-67100 L'Aquila, Italy}
\affiliation{INFN, Laboratori Nazionali del Gran Sasso, I-67100 Assergi, Italy}
\author{Michele Maggiore}
\affiliation{D\'epartement de Physique Th\'eorique and Center for Astroparticle Physics,\\
Universit\'e de Gen\`eve, 24 quai Ansermet, CH--1211 Gen\`eve 4, Switzerland}
\author{B. S. Sathyaprakash}
\penncosmos \pennastro \cardiff
\author{Salvatore Vitale}
\affiliation{LIGO, Massachusetts Institute of Technology, Cambridge, Massachusetts 02139, USA}
\affiliation{Kavli Institute for Astrophysics and Space Research, Massachusetts Institute of Technology, Cambridge, Massachusetts 02139, USA}
\affiliation{Department of Physics, Massachusetts Institute of Technology, Cambridge, Massachusetts 02139, USA}

\date{\today}
\begin{abstract}
The existence of primordial black holes (PBHs), which may form from the collapse of matter overdensities shortly after the Big Bang, is still under debate. Among the potential signatures of PBHs are  gravitational waves (GWs) emitted from binary black hole (BBH) mergers at redshifts $z\gtrsim 30$, where the formation of astrophysical black holes is unlikely.
Future ground-based GW detectors, Cosmic Explorer and Einstein Telescope, will be able to observe equal-mass BBH mergers with total mass of $\mathcal{O}(10-100)~\msun$ at such distances. In this work, we investigate whether the redshift measurement of a single BBH source can be precise enough to establish its primordial origin. We simulate BBHs of different masses, mass ratios and orbital orientations.
We show that for BBHs with total masses between $20~\msun$ and $40~\msun$ merging at $z \geq 40$ one can infer $z>\zcritvalue$ at up to 97\% credibility, with a network of one Einstein Telescope, one 40-km Cosmic Explorer in the US and one 20-km Cosmic Explorer in Australia. A smaller network made of  one Einstein Telescope and one 40-km Cosmic Explorer in the US measures $z>\zcritvalue$ at larger than 90\% credibility for roughly half of the sources than the larger network.
We then assess the dependence of this result on the Bayesian redshift priors used for the analysis, specifically on the relative abundance of the BBH mergers originated from the first stars, and the primordial BBH mergers.
\end{abstract}

\maketitle

\prlsection{Introduction} Formation of primordial black holes (PBHs) was suggested more than five decades ago~\cite{pbh1966zeldovich,Carr:1974nx,Hawking:1971ei}, but these hypothetical compact objects still elude discovery. 
Unlike astrophysical black holes (ABHs) -- which are stellar remnants -- PBHs are formed by the direct collapse of matter overdensities in the early Universe~\cite{Ivanov:1994pa,Garcia-Bellido:1996mdl,Ivanov:1997ia} (also see Refs.~\cite{Polnarev:1985btg,Khlopov:2008qy,Sasaki:2016jop,Green:2020jor} for reviews). While there are significant uncertainties in their mass spectrum, PBH may occur in the range $\mathcal{O}(1$--$100)~\msun$~\cite{Carr:1974nx,Carr:2017jsz,AliHaimoud:2017rtz}.
Stellar-mass PBHs could form binaries~\cite{Nakamura:1997sm,Ioka:1998nz} that merge and emit gravitational waves (GWs) detectable by current GW detectors, including LIGO, Virgo and KAGRA (LVK)~\cite{TheLIGOScientific:2014jea,TheVirgo:2014hva,Aso:2013eba}, and
leave a unique imprint on the mass spectrum and redshift evolution in the observable population of binary black hole (BBH) mergers~\cite{Clesse:2016vqa,Raidal:2017mfl,AliHaimoud:2017rtz,Belotsky:2018wph,Chen:2018czv,Raidal:2018bbj,DeLuca:2020fpg,DeLuca:2020bjf}.
Efforts have been made to test if a fraction of the BBHs in the LVK's second catalog~\cite{GWTC2} could be of primordial origin, by analysing the population properties of BBH mergers detected thus far~\cite{Wong:2020yig,DeLuca:2020qqa,Hutsi:2020sol,Franciolini:2021tla,Mukherjee:2021ags}.
However, even at design sensitivity, the horizon of current GW detectors will be limited to redshifts of $z\lesssim 3$ at most~\cite{Hall:2019xmm}. Interpreting these ``local'' observations, with the aim of establishing the presence of a PBH sub-population, requires precise knowledge of the ABH population, which is dominant at low redshifts and acts as an ``astrophysical foreground''~\cite{Franciolini:2021tla}.
This is challenging, as there exist significant uncertainties on the properties of BBHs formed in different astrophysical environments, such as galactic fields~\cite{OShaughnessy:2016nny,Dominik:2012kk,Dominik:2013tma,Dominik:2014yma,deMink:2015yea,Belczynski:2016obo,Stevenson:2017tfq,Mapelli:2019bnp,Breivik:2019lmt,Bavera:2019fkg,Broekgaarden:2019qnw}, dense star clusters~\cite{2000ApJ...528L..17P,Antonini:2020xnd,Santoliquido:2020axb,Rodriguez:2015oxa,Rodriguez:2016kxx,Rodriguez:2018rmd,DiCarlo:2019pmf,Kremer:2020wtp,Rodriguez:2015oxa,Rodriguez:2016kxx,Rodriguez:2018rmd,Antonini:2020xnd}, active galactic nuclei~\cite{Bartos:2016dgn,Yi:2019rwo,Yang:2019cbr,Yang:2020lhq,Grobner:2020drr,Tagawa:2019osr,Tagawa:2020qll,Tagawa:2020dxe,Samsing:2020tda}, or from the collapse of Population III (Pop III) stars~\cite{Kinugawa:2014zha,Kinugawa:2015nla,Hartwig:2016nde,Belczynski:2016ieo}.

One can ascertain the primordial origin of black holes by detecting mergers at redshifts so high that ABHs could not have had the time to form and merge yet. A plausible lower bound for this redshift would be $z \sim \zcritvalue$ (see discussion below). 
Therefore, measuring the redshift of a BBH merging at redshift larger than \zcritvalue would be a clear hint of the existence of PBHs~\cite{DeLuca:2021hde,DeLuca:2021wjr}.

Proposed next-generation, ground-based GW detectors such as the Cosmic Explorer (CE)~\cite{Evans:2016mbw,Reitze:2019iox,CEHS} and the Einstein Telescope (ET)~\cite{Punturo:2010zz,Maggiore:2019uih} can detect BBH mergers at $z>10$ and above~\cite{Hall:2019xmm}.
However, being able to \emph{detect} a source merging at redshift larger than \zcritvalue does not automatically imply being able to prove that the true redshift was above some threshold.  The purpose of this \letter is to systematically study how well next-generation GW detectors can measure the redshift of distant BBHs.  \\
    
\prlsection{Simulations} Given the significant uncertainty in the mass spectrum of PBHs, we consider a range of values which would lead to detectable GW emission. We simulate BBHs merging at redshifts of $z=10, 20, 30, 40$ and $50$. The total mass in the source frame is chosen to be $\Mtotal=5$, 10, 20, 40, 80, 160, and 250 $\msun$, with mass ratio $q=1$, 2, 3, 4 and 5 (where $q \equiv m_1/m_2$ for $m_1>m_2$) and four values of the orbital inclination $\iota=0$ (face-on), $\pi/6$, $\pi/3$, and $\pi/2$ (edge-on). The simulated BBHs are non-spinning, as it is expected that PBHs are born with negligible spins~\cite{Mirbabayi:2019uph,DeLuca:2019buf} and may be spun-up by accreting materials later in their lives~\cite{Bianchi:2018ula, DeLuca:2020bjf, DeLuca:2020fpg, DeLuca:2020qqa}. However, we do \emph{not} assume that spins are zero when estimating the source parameters and instead allow for generic spin-precession. For each of these 700 sources, the sky location and polarization angles are chosen to maximize the source's signal-to-noise ratio (SNR).

In order to measure a source's distance (and hence redshift~\footnote{Throughout this study, we use a $\Lambda$CDM cosmology based on the Planck 2018 results~\cite{Planck:2018vyg} to convert luminosity distance into redshift.}), one needs to disentangle the two GW polarizations.
Thus, we only consider \emph{networks} of non-collocated next-generation observatories as opposed to single-site detectors. 
We work with two different detector networks: (i) a 40-km CE in the United States and an ET in Europe (CE-ET), and (ii) CE-ET with an additional 20-km CE in Australia (CE-CES-ET).
We only analyze sources whose network SNRs are larger than 12.
Generally speaking, BBHs with $q\geq 5$, $\Mtotal\geq160\msun$ or $\Mtotal\leq5\msun$ can only be detected up to $z\sim20$.

We obtain posterior probability densities with \textsc{bilby}
~\cite{Skilling:2006gxv,2020MNRAS.493.3132S,Ashton:2018jfp}.
The inference is made with a zero-noise realization~\cite{Vallisneri:2007ev}, since we are only interested in the uncertainty caused by limited SNR, and do not want to be prone to offsets potentially caused by large Gaussian fluctuations~\cite{Rodriguez:2013oaa}.
We employ the \texttt{IMRPhenomXPHM} waveform family, which accounts for the effects of spin precession and higher-order modes (HoMs)~\cite{Pratten:2020ceb}, both to create the simulated waveforms and to calculate the likelihood in the sampler. 
It is necessary to include HoMs in our analysis as they are important for systems with large detector-frame masses ($\gtrsim 100\msun$) and large mass ratios. We expect HoMs to be particularly relevant for the analyses we present, as they help break the distance-inclination degeneracy characteristic of the dominant (2,2) multipole mode~\cite{Usman:2018imj,Chen:2018omi}. We find it more efficient to sample the parameter space with uniform priors in the detector-frame total mass, $\Mtotal(1+z)$, and the mass ratio, $q$. The posteriors obtained this way are then reweighed into uniform priors in the source-frame primary mass, $m_1$, and the inverse mass ratio $1/q$ (which is between 0 and 1).
The default prior on redshift is uniform in the source-frame differential comoving volume, $p_0(z) \propto \frac{dV_c}{dz}\frac{1}{1+z}$, but we will explore other options below.
We used isotropic priors for the sky position, the orbital and spin orientations. Uniform priors were used for the arrival time, phase of the signal at the time of arrival, and the spin magnitude.
Unless otherwise specified, in what follows we quote uncertainties at 95\% credible intervals.\\

\begin{figure}[ht]
    \centering
    \includegraphics[width=0.9\columnwidth]{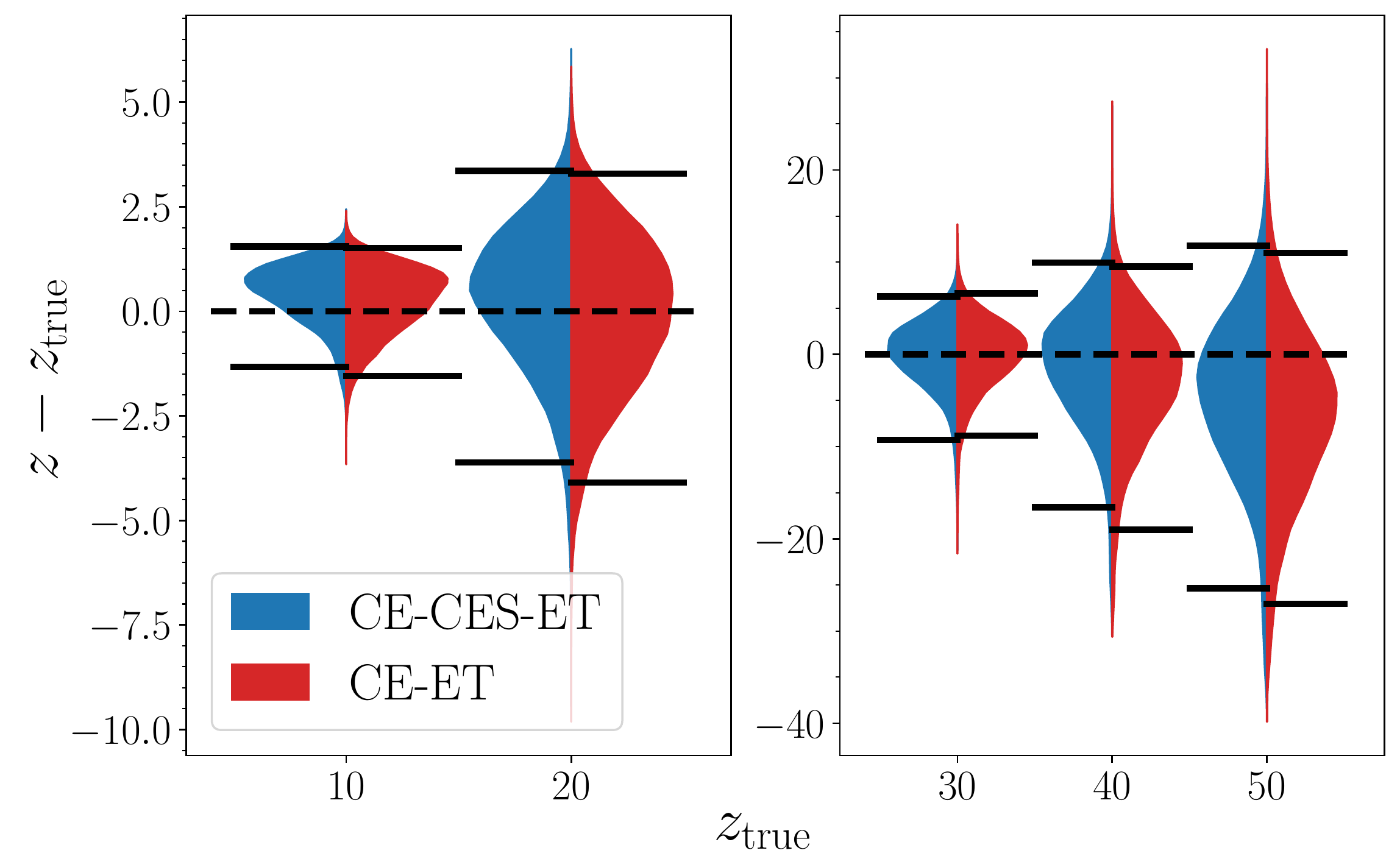}
    \caption{Redshift posteriors (offset by the true values) for sources with $(\Mtotal,q,\iota)=(40\msun,1,\pi/6)$ at 5 different redshifts observed by CE-ET (red) and CE-CES-ET (blue). The solid horizontal lines show the 95\% credible intervals, whereas the dashed lines mark \ztrue.}
    \label{fig:posVSz}
\end{figure}

\prlsection{Redshift uncertainties} Figure~\ref{fig:posVSz} shows the redshift posteriors for a BBH system with $(\Mtotal,q,\iota)=(40\msun,1,\pi/6)$ located at different redshifts (given in the $x$-axis),  as observed by CE-ET (red) and CE-CES-ET (blue). To increase the resolution along the $y$-axis, we have offset each posterior by the true redshift value.
The redshift uncertainty increases with the true redshift, from $\Delta z \sim 5$ at $\ztrue=10$ to $\Delta z\sim 30$ at $\ztrue=50$ for a CE-CES-ET network.

The roughly linear increase of the uncertainty with redshift is not just due to the reduction of the signal's amplitude, as it happens for the sources discovered by advanced detectors. For any frequency in the inspiral phase of the waveform, the Fourier amplitude is proportional to
${\left[(1+z)\Mc \right]^{5/6}}{d_L}^{-1}$, where $\Mc$ is the source-frame chirp mass and $d_L$ is the luminosity distance. At large redshifts $d_L\propto (1+z).$ Thus, for a given source-frame $\Mc$, the inspiral Fourier amplitude of sources at $z\gg1$ only falls off as $(1+z)^{-1/6}$. However, the loss in SNR at higher redshifts is greater because frequencies are redshifted by a factor of $(1+z)$ where the detectors' sensitivity might be poorer but also because signals have smaller bandwidth in the detectors' sensitivity range.
These effects add up to yield the trend observed in Fig.~\ref{fig:posVSz}.

We also note that the network without CES yields uncertainties which are up to $\sim 10\%$ larger than the three-detector network.
This happens because an extra detector improves the resolution of the GW polarizations.

\begin{figure}[h]
    \centering
    \includegraphics[width=0.9\columnwidth]{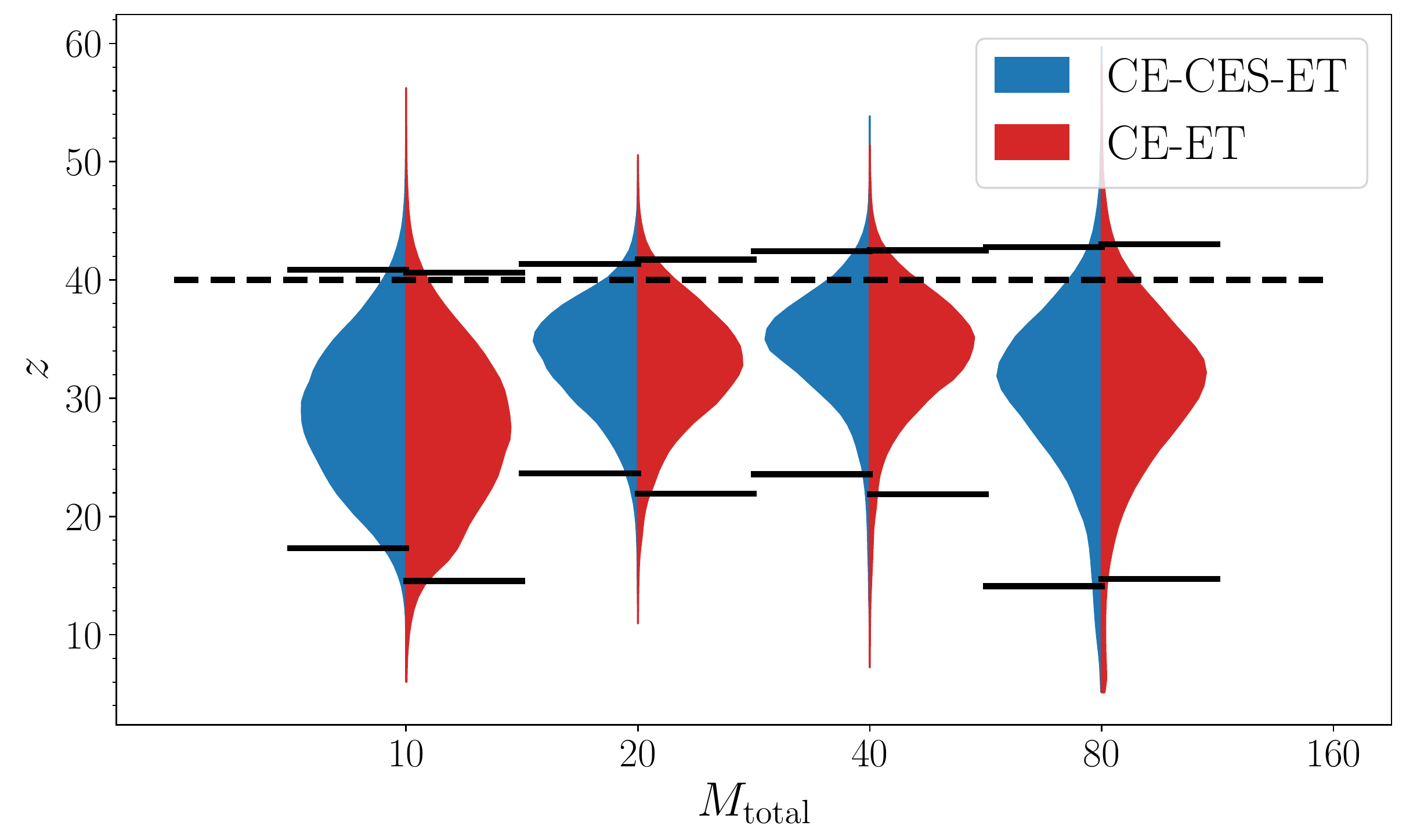}
    \caption{Redshift posteriors for sources with $(q,\iota,z)=(1,0,40)$ at 4 different total masses observed by CE-ET (red) and CE-CES-ET (blue).
    The solid horizontal lines show the 95\% credible intervals, whereas the dashed line marks \ztrue.
    Only sources with $\rm SNR\geq 12$ are included.
    }
    \label{fig:posVSmtotal}
\end{figure}

Next, we discuss how the redshift uncertainty depends on the total mass of the system. In Fig.~\ref{fig:posVSmtotal} we show the redshift posteriors of BBHs of increasing $\Mtotal$, with $\ztrue = 40$,  $q=1$ and $\iota=0$.
The redshift uncertainty obtained by CE-CES-ET first decreases from $\Delta z \sim 20$ for $\Mtotal=10 \msun$ to $\Delta z \sim 15$ for $\Mtotal=40 \msun$, then increases to $\Delta z\sim 25$ for $\Mtotal=80 \msun$ (this plot does not feature a system with $\Mtotal=5$ and $160~\msun$ as its SNR is below the threshold of 12). The non-linear trend apparent in this plot is the result of two competing effects. At first, as the mass increases, so do the signal amplitude and the SNR. In this regime, the uncertainty decreases with $\Mtotal$. However, when $\Mtotal$ increases further, signals sweep less of the detectors' sensitive bandwidth. In this regime, the uncertainties increase with $\Mtotal$.

We also notice an offset between the true redshift and the maximum \textit{a posteriori} redshift of each source.
There are two reasons for this offset.
The default redshift prior goes as $p_0(z) \sim (1+z)^{-5/2}$ in the matter-dominated regime, $1\lesssim z\lesssim 1000$, hence favoring lower redshifts.
Furthermore, the true inclination of these sources is $0$, which implies that underestimating the redshift can be compensated for by measuring inclination angles closer to edge-on. The opposite is obviously not possible since the true inclination is at the edge of its physically allowed prior.
This asymmetry makes it easier to underestimate the redshift but more difficult to overestimate it. 
This is why this systematic offset was not observed in Fig.~\ref{fig:posVSz}, where the true inclination was $\pi/6$.
The effect of priors is discussed more extensively below.\\

\prlsection{Can single-event-based PBH identification be unambiguous?} With these results at hand, we now address the key question of this study: can one pinpoint the redshift of merging BBHs to be above some threshold value? The answer will clearly depend on which threshold is used. 
While the redshift at which the first stars were born is not precisely known, theoretical calculations and cosmological simulations suggest that the first stars might have been born at redshifts smaller than $z\sim 40$~\cite{Bromm:2005ep,deSouza:2011ea,Koushiappas:2017kqm,Mocz:2019uyd}~\footnote{Although, we note that there are studies suggesting an earlier $(z\gtrsim 50)$~\cite{Trenti:2009cj} or a later $(z\lesssim20)$~\cite{Tornatore:2007ds} formation of Pop~III stars.}.
Meanwhile, population synthesis models indicate that the characteristic time delay between formation and merger for Pop~III stars is $\mathcal{O}(10)~\rm Myr$~\cite{Kinugawa:2014zha,Kinugawa:2015nla,Hartwig:2016nde,Belczynski:2016ieo,Inayoshi:2017mrs,Liu:2020lmi,Liu:2020ufc,Kinugawa:2020ego,Tanikawa:2020cca}.
This implies that the redshift at which Pop~III remnant BBHs merge is below $z\sim 30$.
On the other hand, {stellar-mass} PBHs are expected to have formed {at $z\gg 1000$, much earlier than} the recombination epoch, and have been merging with one another since then~\cite{Raidal:2018bbj}.
The PBH merger rate density therefore increases monotonically with redshift~\cite{Raidal:2018bbj}, unlike that of ABHs.

Based on the above, we choose a critical redshift $\zcrit=\zcritvalue$, above which no astrophysical BBHs are expected to merge.
We define the probability of primordial origin, $P_\text{p}$ as the fraction of the redshift posterior with $z\geq \zcrit$.
In the following, we only consider the measurements made by CE-CES-ET which are always better than those made by CE-ET.

\input{table_Pp_zcrit30}

In Table~\ref{tab:posMasszcrit30}, we list the BBH sources for which $P_\text{p}\geq0.9$, and the corresponding values of $P_\text{p}$.
We find that only sources with $z\geq40$ and $\Mtotal=20\msun$ or $40\msun$ achieve $P_\text{p}\geq0.9$.
We stress that the highest $P_\text{p}$'s are not reached by face-on sources, despite the fact that they have the largest SNRs. This is because the amplitude of HoMs is smallest for face-on sources, which will suffer the most from the distance-inclination degeneracy, and hence have larger uncertainties.
We will explore this trade-off further in a forthcoming paper.

Next, we want to explore the dependence of these results on the redshift prior. The results presented thus far used our default prior for the redshift, uniform in co-moving volume and source-frame time, 
given above. One could instead decide to use a prior informed by reasonable merger rates for ABHs and PBHs. In this case, the analysis of $P_\text{p}$ answers the following question: given some expected distribution of merger rates of PBHs and ABHs, what is the probability that \emph{this} system is primordial?
To answer quantitatively, we adopt for the PBHs the merger rate density $\npbh$, from Refs.~\cite{Raidal:2017mfl,Raidal:2018bbj,DeLuca:2020qqa} $\npbh(z) \propto \left(\frac{t(z)}{t(0)}\right)^{-34/37}$, and that for the Pop~III BBH mergers, $\nIII$, from Refs.~\cite{Belczynski:2016ieo,Ng:2020qpk},
\begin{align}\label{eq:popIII}
    \nIII(z) \propto
    \begin{cases}
     \frac{e^{\aIII(z-\zIII)}}{\bIII+\aIII e^{(\aIII+\bIII)(z-\zIII)}} &\mathrm{if }\,z<\zcrit \\
     0 &\mathrm{otherwise}
    \end{cases},
\end{align}
with $t(z)$ the age of the Universe at $z$, and $(\aIII,\bIII,\zIII)=(0.66, 0.3, 11.6)$~\cite{Ng:2020qpk}.

Then, we construct a mixture model for the merger rate based redshift prior, $\ptot(z|\fcrit)$,
\begin{align}
    \ptot(z|\fcrit) \propto \left[\fcrit\frac{\npbh(z)}{\npbh(\zcrit)}+\frac{\nIII(z)}{\nIII(\zcrit)}\right]\frac{dV_c}{dz}\frac{1}{1+z},\nonumber
\end{align}
where $\fcrit$ represents the ratio of the merger rate contributed by PBH mergers to that contributed by Pop~III BBH mergers at $\zcrit$: $\fcrit\equiv {\npbh(\zcrit)}/{\nIII(\zcrit)}$.
We plot $\ptot(z|\fcrit)$ for 5 different value of $\fcrit$ in the \supmat. 

\begin{figure}[h]
    \centering
    \includegraphics[width=0.9\columnwidth]{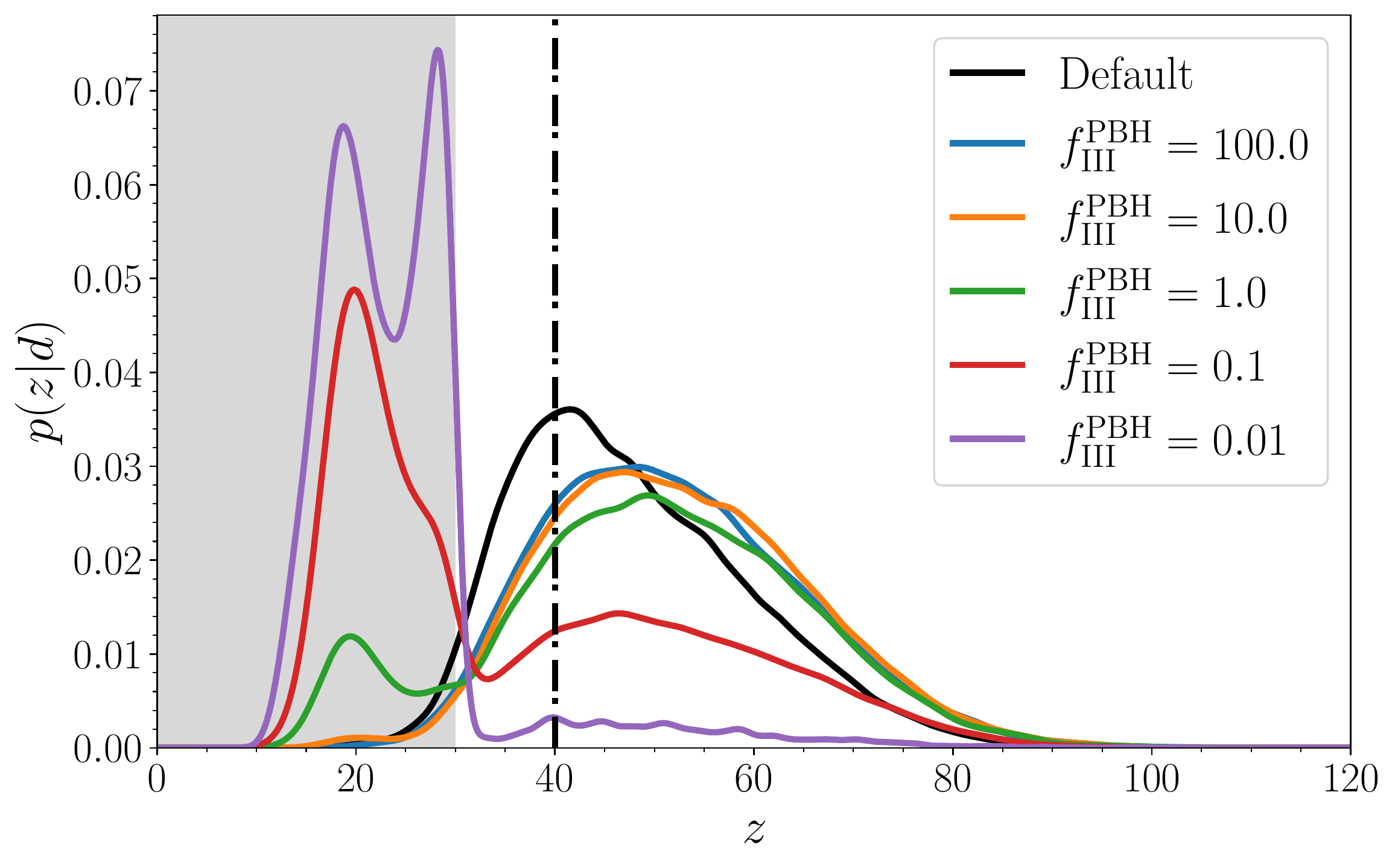}
    \caption{Posteriors of the system $(\Mtotal,q,\iota,z)=(40\msun,1,\pi/3,40)$, using $\ptot(z)$ evaluated at $\fcrit=100$ (blue), 10 (orange), 1 (green), 0.1 (red), and 0.01 (purple).
    The black solid line shows the default posterior.
    The black dashed line indicates $\ztrue=40$, and the grey area indicates the astrophysical region $z<\zcrit$.
    }
    \label{fig:posVSfcrit}
\end{figure}

We then re-analyze the source with the largest $P_\text{p}$ of Table~\ref{tab:posMasszcrit30}, namely, $(\Mtotal,q,\iota,z)=(40\msun,1,\pi/3,40)$, by applying the rate-based redshift prior $\ptot(z|\fcrit)$ instead of the default prior $p_0(z)$.
The redshift posteriors for different $\fcrit$ are shown in Fig.~\ref{fig:posVSfcrit}.
When $\fcrit$ decreases below 1, the posterior becomes bimodal with a low-redshift peak around $z\approx 20$, which becomes the dominating peak for $\fcrit\lesssim0.1$.
We tabulate the values of $P_\text{p}$ 
for the same system, calculated with the different priors in Table~\ref{tab:posMassReweigh}.
The value of $P_\text{p}$ drops below $0.95$ for $\fcrit\leq 1$.
This seems to suggest that unless one believes \textit{a priori} that the relative fraction between PBH mergers and Pop~III BBH mergers at \zcrit is at least unity, the probability that a specific source is primordial will be low.

However, not all priors yield comparable Bayesian evidences. One can thus calculate the Bayes factor  $\BF = Z_{\rm P}/Z_0$, where $Z_{\rm P}$ and $Z_0$ are the single-event evidences calculated with the rate-based prior and the default prior, respectively~\cite{Vitale:2017cfs,Zevin:2020gxf,Bhagwat:2020bzh}.
Values of $\BF>1$ imply that the default prior is disfavoured.
As shown in Table~\ref{tab:posMassReweigh}, the value of $\BF$ decreases from $\sim2$ to $\sim 3\times 10^{-3}$ with $\fcrit$, meaning that rate-based priors with small PBH fraction $\fcrit\lesssim0.1$ are strongly disfavored when comparing to the rate based prior with $\fcrit\gtrsim 1$.
Therefore, while an \textit{a-priori} belief that the fraction of PBHs is low will move the redshift posterior to lower values, it will also yield a low Bayes factor that disfavors the model with a smaller $\fcrit$.

\input{table_bayes_factor}

\prlsection{Discussion} In this \letter, we have simulated BBHs merging at $z\geq10$ and quantified the uncertainties in their redshifts as measured by networks of next-generation ground-based gravitational-wave detectors. 
We have used full Bayesian parameter estimation, and a waveform model which accounts for both higher-order modes and spin precession.
The relative redshift uncertainties are larger than $\sim10\%$ at $z\geq10$, even for a three-detector network CE-CES-ET.
Assuming that no astrophysical BBHs can  merge above $\zcrit=\zcritvalue$ (see \supmat for a lower threshold), we found that the typical redshift measurement is not precise enough to conclude with certainty that a single source is of primordial origin.
Among the systems we simulated, the ones for which one can establish the primordial nature more strongly have $\Mtotal=20\msun$ or $40\msun$ at $\ztrue \geq 40$.
We find 12 such systems for which $90\% <p(z>\zcrit)<97\%$ assuming a uniform prior in source-frame differential comoving volume and using a CE-CES-ET network.
With a smaller CE-ET network, only 3 sources have $90\% <p(z>\zcrit)$. We have also verified that with a single ET observatory one can reach $p(z>\zcrit)\sim 70\%$. With a smaller set of targeted simulations, we have verified that with a single CE observatory one cannot set significant constraints. This is expected as the capabilities of a single L-shaped detector to measure distances outside of a network are limited.

Next, we have shown how the redshift measurement depends significantly on the redshift prior used for the analysis. In particular, we have considered a prior informed by population synthesis and theoretical models and introduced a mixture model that allows for both ABHs and PBHs. As one would expect, the posterior probability of primordial origin for a specific event decreases if the assumed prior ratio of PBH merger rate to Pop~III BBH merger rate is small. However, the Bayes factor between the mixture model and the constant comoving density model also decreases.
Claims based on individual events would thus benefit from better knowledge of the properties of Pop~III stars and their remnants, notably the highest rate at which they merge. Forthcoming facilities such as the JWST, the Roman space telescope or Euclid are expected to probe the properties of Pop~III stars by accessing Pop~III galaxies in blind surveys~\cite{Vikaeus2021}. More information will be yielded by instrument such as SPHEREx, by precisely measuring the near infrared background \cite{Sun2021}; and mission concepts like THESEUS, by detecting the most distant long gamma-ray burst \cite{Tanvir2021}.

While an event-by-event basis identification of PBH binaries might be challenging, one can  find evidence for PBHs in the whole dataset.
There are two possible methods to perform such an analysis.
First, the subthreshold events from PBH mergers may contribute to a distinctive stochastic GW background, whose amplitude is directly related to the rate of PBHs  mergers~\cite{Mukherjee:2021itf}.
Second, one may use multiple resolvable sources at high redshifts and combine their redshift measurements through hierarchical Bayesian analysis. This also allows for detailed modelling on the features of each subpopulation~\cite{Farr:2013yna,Wysocki:2018mpo,Thrane:2018qnx,Mandel:2018mve,Vitale:2020aaz}. We will explore this avenue in a future paper.

\prlsection{Acknowledgments} 
We would like to thank E. Berti, M. Evans, G. Franciolini, P. Pani, M. Punturo, A. Riotto for suggestions and comments. 
KKYN and SV are supported by the NSF through the award PHY-1836814. SC is supported by the Undergraduate Research Opportunities Program of Massachusetts Institute of Technology.
The work  of MM is supported by the  Swiss National Science Foundation and  by the SwissMap National Center for Competence in Research. MB acknowledges support from the European Union’s Horizon 2020 Programme under the AHEAD2020 project (grant agreement n. 871158). BSS is supported in part by NSF Grant No. PHY-1836779, AST-2006384 and PHY- 2012083. SB is supported by NSF Grant No.PHY-1836779. BG is supported by the Italian Ministry of Education, University and Research within the PRIN 2017 Research Program Framework, n. 2017SYRTCN.

\bibliography{pbh.bib}
\clearpage

\begin{figure}[h!]
    \centering
    \includegraphics[width=0.9\columnwidth]{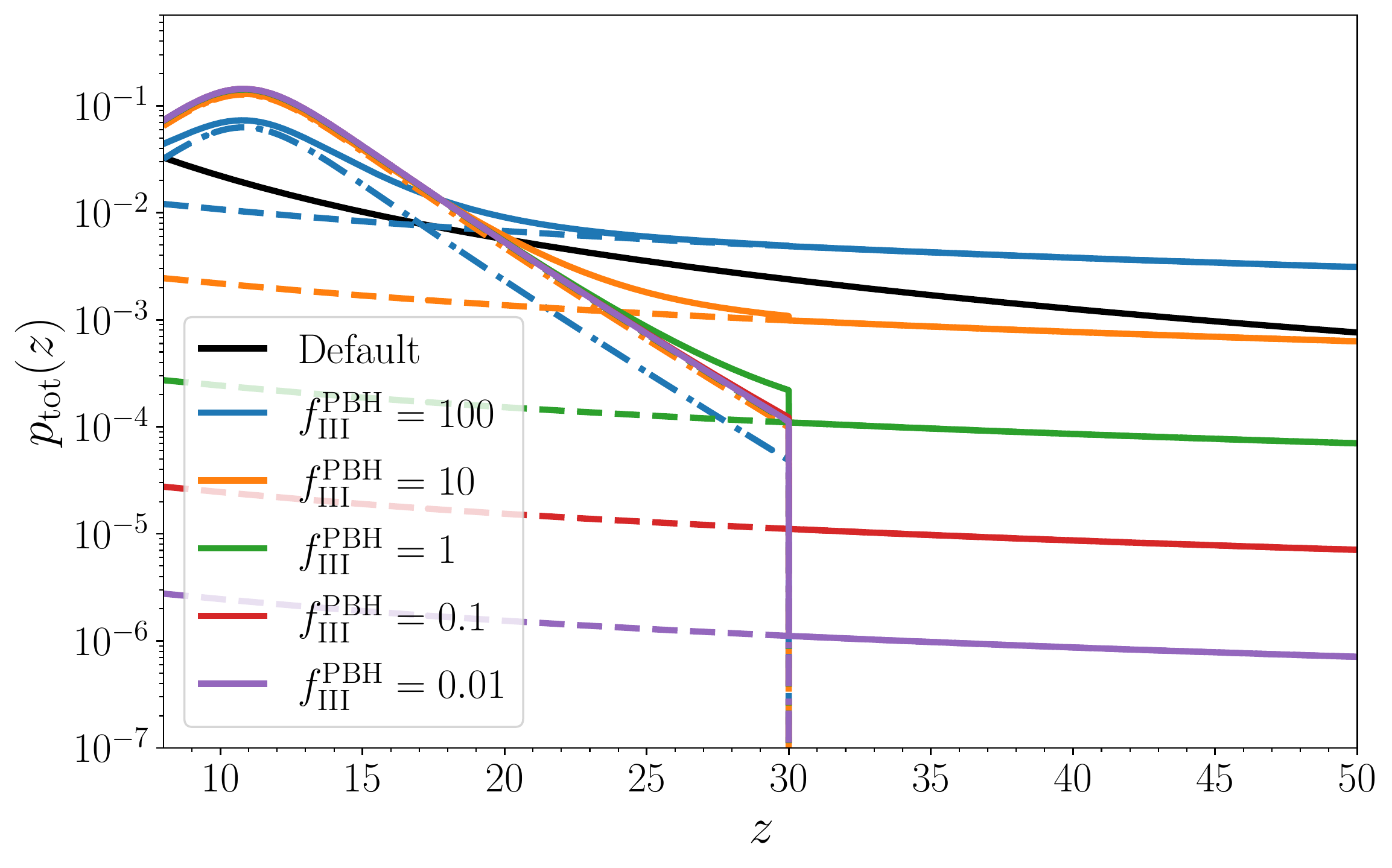}
    \caption{Prior based on the mixture model $\ptot$ (solid lines), ranging from $\fcrit=100$ (blue), 10 (orange), 1 (green), 0.1 (red), and 0.01 (purple).
    The dashed lines and dashed-dotted lines represent the individual contribution from PBH mergers and Pop~III BBH mergers (which is truncated at $\zcrit$), respectively.
    }
    \label{fig:ptot_z}
\end{figure}

\prlsection{Supplementary Materials}
In Fig.~\ref{fig:ptot_z}, we show $\ptot(z|\fcrit)$ for 5 different values of $\fcrit$.
The appearance of the discontinuity at $\zcrit$ is caused by the piecewise nature of the Pop~III BBH merger rate density in Eq.~(3) of the \letter.

Given that the peak of the merger rate for BBHs from Pop~III stars is at $z\sim 12$ (Fig.~\ref{fig:ptot_z}), one might explore values of $\zcrit$ lower than what we used in the body of the \letter. Below, we report a version of Table~I obtained with a less conservative $\zcrit=20$. This result in a larger number of sources that clear the $P_p>0.9$ criterion, and even yields sources for which all of the posterior is above \zcrit.

\squeezetable
\begin{table*}[h!]
    \centering
    \caption{Same as Table~I but with $\zcrit=20$}
    \begin{ruledtabular}
    \begin{tabular}{c c c c c c c c c}
$\Mtotal$ & $q$ & $\iota$ & $z$ & $\rho_{\rm CE}$ & $\rho_{\rm CES}$ & $\rho_{\rm ET}$ & $\rho_{\rm net}$ & $P_p$ \\
\hline
10 & 1 & 0 &40 & 12.1 & 5.8 & 3.7 & 13.9 &  0.92 \\
10 & 1 & 0 &50 & 11.3 & 5.4 & 3.4 & 13.0 &  0.98 \\
20 & 1 & 0 &30 & 22.7 & 10.6 & 7.0 & 26.0 &  0.96 \\
20 & 1 & 0 &40 & 19.9 & 8.5 & 6.0 & 22.5 &  1.00 \\
20 & 1 & 0 &50 & 16.6 & 7.2 & 5.9 & 19.1 &  0.99 \\
20 & 2 & 0 &30 & 19.4 & 9.0 & 6.2 & 22.3 &  0.99 \\
20 & 2 & 0 &40 & 16.5 & 7.3 & 5.5 & 18.9 &  0.99 \\
20 & 2 & 0 &50 & 13.6 & 5.8 & 5.2 & 15.7 &  0.99 \\
20 & 3 & 0 &30 & 15.7 & 7.2 & 5.1 & 18.0 &  0.98 \\
20 & 3 & 0 &40 & 13.0 & 5.7 & 4.5 & 14.9 &  1.00 \\
20 & 4 & 0 &30 & 13.0 & 5.8 & 4.4 & 14.9 &  0.98 \\
20 & 5 & 0 &30 & 11.0 & 4.9 & 3.8 & 12.6 &  0.99 \\
20 & 1 & $\pi/6$ & 30 & 19.7 & 8.7 & 5.7 & 22.3 &  0.96 \\
20 & 1 & $\pi/6$ & 40 & 17.0 & 7.3 & 5.1 & 19.2 &  0.99 \\
20 & 1 & $\pi/6$ & 50 & 14.0 & 6.3 & 5.2 & 16.2 &  1.00 \\
20 & 2 & $\pi/6$ & 30 & 17.3 & 7.6 & 5.0 & 19.6 &  0.95 \\
20 & 2 & $\pi/6$ & 40 & 14.2 & 6.1 & 4.4 & 16.1 &  0.99 \\
20 & 2 & $\pi/6$ & 50 & 11.6 & 5.1 & 4.5 & 13.5 &  0.99 \\
20 & 3 & $\pi/6$ & 30 & 14.3 & 6.2 & 4.3 & 16.1 &  0.97 \\
20 & 3 & $\pi/6$ & 40 & 11.1 & 5.0 & 4.0 & 12.8 &  0.99 \\
20 & 4 & $\pi/6$ & 30 & 11.7 & 5.5 & 4.1 & 13.5 &  0.97 \\
20 & 1 & $\pi/3$ & 30 & 13.1 & 5.7 & 3.3 & 14.7 &  0.98 \\
20 & 1 & $\pi/3$ & 40 & 11.2 & 4.7 & 3.0 & 12.5 &  1.00 \\
20 & 2 & $\pi/3$ & 30 & 12.2 & 5.4 & 3.0 & 13.7 &  0.95 \\
40 & 1 & 0 &30 & 27.6 & 12.6 & 13.3 & 33.1 &  0.98 \\
40 & 1 & 0 &40 & 8.8 & 8.2 & 19.8 & 23.2 &  0.99 \\
40 & 1 & 0 &50 & 3.0 & 4.2 & 17.8 & 18.5 &  0.99 \\
40 & 2 & 0 &30 & 21.5 & 10.1 & 12.2 & 26.8 &  0.98 \\
40 & 2 & 0 &40 & 5.6 & 6.0 & 17.2 & 19.0 &  0.98 \\
40 & 2 & 0 &50 & 2.1 & 3.0 & 14.5 & 15.0 &  0.98 \\
40 & 3 & 0 &30 & 14.6 & 7.8 & 11.5 & 20.2 &  0.96 \\
40 & 3 & 0 &40 & 3.2 & 4.0 & 13.7 & 14.7 &  0.98 \\
40 & 4 & 0 &30 & 10.1 & 6.2 & 10.7 & 16.0 &  0.97 \\
40 & 5 & 0 &30 & 7.6 & 5.0 & 9.5 & 13.2 &  0.99 \\
40 & 1 & $\pi/6$ & 30 & 23.0 & 10.8 & 12.0 & 28.1 &  0.98 \\
40 & 1 & $\pi/6$ & 40 & 7.4 & 6.9 & 16.8 & 19.6 &  0.99 \\
40 & 1 & $\pi/6$ & 50 & 2.6 & 3.5 & 15.1 & 15.7 &  0.99 \\
40 & 2 & $\pi/6$ & 30 & 19.2 & 9.0 & 10.5 & 23.7 &  0.93 \\
40 & 2 & $\pi/6$ & 40 & 5.2 & 5.6 & 14.9 & 16.8 &  0.96 \\
40 & 2 & $\pi/6$ & 50 & 2.2 & 3.0 & 12.8 & 13.3 &  0.98 \\
40 & 3 & $\pi/6$ & 40 & 3.4 & 4.2 & 12.2 & 13.4 &  0.96 \\
40 & 4 & $\pi/6$ & 30 & 10.2 & 6.3 & 9.4 & 15.2 &  0.92 \\
40 & 1 & $\pi/3$ & 30 & 15.3 & 6.9 & 6.6 & 18.0 &  1.00 \\
40 & 1 & $\pi/3$ & 40 & 6.0 & 4.6 & 9.5 & 12.1 &  1.00 \\
40 & 2 & $\pi/3$ & 30 & 13.8 & 6.3 & 5.9 & 16.3 &  0.97 \\
40 & 1 & $\pi/2$ & 30 & 11.5 & 4.6 & 2.8 & 12.7 &  1.00 \\
80 & 1 & 0 &30 & 2.7 & 4.5 & 29.1 & 29.5 &  0.98 \\
80 & 1 & 0 &40 & 0.8 & 1.4 & 14.1 & 14.2 &  0.93 \\
80 & 2 & 0 &30 & 1.9 & 3.2 & 22.4 & 22.7 &  0.97 \\
80 & 3 & 0 &30 & 1.2 & 2.0 & 15.7 & 15.9 &  0.94 \\
80 & 1 & $\pi/6$ & 30 & 2.3 & 3.8 & 24.6 & 25.0 &  0.99 \\
80 & 1 & $\pi/3$ & 30 & 1.3 & 2.7 & 14.8 & 15.1 &  1.00 \\
80 & 2 & $\pi/3$ & 30 & 1.4 & 3.0 & 13.4 & 13.8 &  0.97 \\
    \end{tabular}
    \end{ruledtabular}
    \label{tab:posMasszcrit20}
\end{table*}

\end{document}

%% file: table_Pp_zcrit30.tex
\begin{table}[h]
    \centering
    \caption{Parameters of the simulations that result in a probability of a primordial origin $P_\text{p}\geq0.9$, assuming the default prior $p_0(z)$.
    We also show the individual SNR in CE, $\rho_{\rm CE}$, SNR in CES, $\rho_{\rm CES}$, SNR in ET, $\rho_{\rm ET}$, as well as the network SNR, $\rho_{\rm net}$.
    }
    \renewcommand{\arraystretch}{1.2} 
    \begin{ruledtabular}
    \begin{tabular}{c c c c c c c c c}
$\Mtotal$ & $q$ & $\iota$ & $z$ & $\rho_{\rm CE}$ & $\rho_{\rm CES}$ & $\rho_{\rm ET}$ & $\rho_{\rm net}$ & $P_p$ \\
\hline
20 & 1 & 0 &50 & 16.6 & 7.2 & 5.9 & 19.1 &  0.93 \\
20 & 1 & $\pi/6$ & 50 & 14.0 & 6.3 & 5.2 & 16.2 &  0.94 \\
20 & 2 & 0 &50 & 13.6 & 5.8 & 5.2 & 15.7 &  0.95 \\
20 & 2 & $\pi/6$ & 50 & 11.6 & 5.1 & 4.5 & 13.5 &  0.91 \\
20 & 3 & 0 &40 & 13.0 & 5.7 & 4.5 & 14.9 &  0.92 \\
40 & 1 & $\pi/6$ & 40 & 7.4 & 6.9 & 16.8 & 19.6 &  0.91 \\
40 & 1 & $\pi/3$ & 40 & 6.0 & 4.6 & 9.5 & 12.1 &  0.97 \\
40 & 1 & 0 &50 & 3.0 & 4.2 & 17.8 & 18.5 &  0.94 \\
40 & 1 & $\pi/6$ & 50 & 2.6 & 3.5 & 15.1 & 15.7 &  0.96 \\
40 & 2 & 0 &40 & 5.6 & 6.0 & 17.2 & 19.0 &  0.92 \\
40 & 2 & 0 &50 & 2.1 & 3.0 & 14.5 & 15.0 &  0.93 \\
40 & 3 & 0 &40 & 3.2 & 4.0 & 13.7 & 14.7 &  0.91 \\
    \end{tabular}
    \end{ruledtabular}
    \label{tab:posMasszcrit30}
\end{table}

%% file: table_bayes_factor.tex
\begin{table}[h]
    \centering
    \caption{Probabilities of a primordial origin $P_\text{p}$ of the system $(\Mtotal,q,\iota,z)=(40\msun,1,\pi/3,40)$ and the Bayes factor $\BF$ against the default prior calculated with the rate based priors $\ptot(z|\fcrit)$ for different PBH-to-Pop III merger rate ratios at $\zcrit$, $\fcrit$.
    The value of $P_{\rm p}$ calculated with default prior $p_0(z)$ is shown for comparison.}
    \renewcommand{\arraystretch}{1.2} 
    \begin{ruledtabular}
    \begin{tabular}{l c c}
    $\fcrit$ & $P_\text{p}$ & $\BF$ \\
    \hline
    100 & 0.98 & 1.93\\
    10 & 0.98 & 0.40 \\
    1 & 0.88 & 0.035 \\
    0.1 & 0.49 & 0.0030 \\
    0.01 & 0.11 & 0.0028 \\
    \hline
    Default & 0.97 & 1\\
    \end{tabular}
    \end{ruledtabular}
    \label{tab:posMassReweigh}
\end{table}

%% file: pbh_pe_prl.bbl
\begin{thebibliography}{108}%
\makeatletter
\providecommand \@ifxundefined [1]{%
 \@ifx{#1\undefined}
}%
\providecommand \@ifnum [1]{%
 \ifnum #1\expandafter \@firstoftwo
 \else \expandafter \@secondoftwo
 \fi
}%
\providecommand \@ifx [1]{%
 \ifx #1\expandafter \@firstoftwo
 \else \expandafter \@secondoftwo
 \fi
}%
\providecommand \natexlab [1]{#1}%
\providecommand \enquote  [1]{``#1''}%
\providecommand \bibnamefont  [1]{#1}%
\providecommand \bibfnamefont [1]{#1}%
\providecommand \citenamefont [1]{#1}%
\providecommand \href@noop [0]{\@secondoftwo}%
\providecommand \href [0]{\begingroup \@sanitize@url \@href}%
\providecommand \@href[1]{\@@startlink{#1}\@@href}%
\providecommand \@@href[1]{\endgroup#1\@@endlink}%
\providecommand \@sanitize@url [0]{\catcode `\\12\catcode `\$12\catcode
  `\&12\catcode `\#12\catcode `\^12\catcode `\_12\catcode `\%12\relax}%
\providecommand \@@startlink[1]{}%
\providecommand \@@endlink[0]{}%
\providecommand \url  [0]{\begingroup\@sanitize@url \@url }%
\providecommand \@url [1]{\endgroup\@href {#1}{\urlprefix }}%
\providecommand \urlprefix  [0]{URL }%
\providecommand \Eprint [0]{\href }%
\providecommand \doibase [0]{https://doi.org/}%
\providecommand \selectlanguage [0]{\@gobble}%
\providecommand \bibinfo  [0]{\@secondoftwo}%
\providecommand \bibfield  [0]{\@secondoftwo}%
\providecommand \translation [1]{[#1]}%
\providecommand \BibitemOpen [0]{}%
\providecommand \bibitemStop [0]{}%
\providecommand \bibitemNoStop [0]{.\EOS\space}%
\providecommand \EOS [0]{\spacefactor3000\relax}%
\providecommand \BibitemShut  [1]{\csname bibitem#1\endcsname}%
\let\auto@bib@innerbib\@empty
\bibitem [{\citenamefont {{Zel'dovich}}\ and\ \citenamefont
  {{Novikov}}(1966)}]{pbh1966zeldovich}%
  \BibitemOpen
  \bibfield  {author} {\bibinfo {author} {\bibfnamefont {Y.~B.}\ \bibnamefont
  {{Zel'dovich}}}\ and\ \bibinfo {author} {\bibfnamefont {I.~D.}\ \bibnamefont
  {{Novikov}}},\ }\bibfield  {title} {\bibinfo {title} {{The Hypothesis of
  Cores Retarded during Expansion and the Hot Cosmological Model}},\
  }\href@noop {} {\bibfield  {journal} {\bibinfo  {journal} {\azh}\ }\textbf
  {\bibinfo {volume} {43}},\ \bibinfo {pages} {758} (\bibinfo {year}
  {1966})}\BibitemShut {NoStop}%
\bibitem [{\citenamefont {Carr}\ and\ \citenamefont
  {Hawking}(1974)}]{Carr:1974nx}%
  \BibitemOpen
  \bibfield  {author} {\bibinfo {author} {\bibfnamefont {B.~J.}\ \bibnamefont
  {Carr}}\ and\ \bibinfo {author} {\bibfnamefont {S.}~\bibnamefont {Hawking}},\
  }\bibfield  {title} {\bibinfo {title} {{Black holes in the early Universe}},\
  }\href@noop {} {\bibfield  {journal} {\bibinfo  {journal} {Mon. Not. Roy.
  Astron. Soc.}\ }\textbf {\bibinfo {volume} {168}},\ \bibinfo {pages} {399}
  (\bibinfo {year} {1974})}\BibitemShut {NoStop}%
\bibitem [{\citenamefont {Hawking}(1971)}]{Hawking:1971ei}%
  \BibitemOpen
  \bibfield  {author} {\bibinfo {author} {\bibfnamefont {S.}~\bibnamefont
  {Hawking}},\ }\bibfield  {title} {\bibinfo {title} {{Gravitationally
  collapsed objects of very low mass}},\ }\href@noop {} {\bibfield  {journal}
  {\bibinfo  {journal} {Mon. Not. Roy. Astron. Soc.}\ }\textbf {\bibinfo
  {volume} {152}},\ \bibinfo {pages} {75} (\bibinfo {year} {1971})}\BibitemShut
  {NoStop}%
\bibitem [{\citenamefont {Ivanov}\ \emph {et~al.}(1994)\citenamefont {Ivanov},
  \citenamefont {Naselsky},\ and\ \citenamefont {Novikov}}]{Ivanov:1994pa}%
  \BibitemOpen
  \bibfield  {author} {\bibinfo {author} {\bibfnamefont {P.}~\bibnamefont
  {Ivanov}}, \bibinfo {author} {\bibfnamefont {P.}~\bibnamefont {Naselsky}},\
  and\ \bibinfo {author} {\bibfnamefont {I.}~\bibnamefont {Novikov}},\
  }\bibfield  {title} {\bibinfo {title} {{Inflation and primordial black holes
  as dark matter}},\ }\href {https://doi.org/10.1103/PhysRevD.50.7173}
  {\bibfield  {journal} {\bibinfo  {journal} {Phys. Rev. D}\ }\textbf {\bibinfo
  {volume} {50}},\ \bibinfo {pages} {7173} (\bibinfo {year}
  {1994})}\BibitemShut {NoStop}%
\bibitem [{\citenamefont {Garcia-Bellido}\ \emph {et~al.}(1996)\citenamefont
  {Garcia-Bellido}, \citenamefont {Linde},\ and\ \citenamefont
  {Wands}}]{Garcia-Bellido:1996mdl}%
  \BibitemOpen
  \bibfield  {author} {\bibinfo {author} {\bibfnamefont {J.}~\bibnamefont
  {Garcia-Bellido}}, \bibinfo {author} {\bibfnamefont {A.~D.}\ \bibnamefont
  {Linde}},\ and\ \bibinfo {author} {\bibfnamefont {D.}~\bibnamefont {Wands}},\
  }\bibfield  {title} {\bibinfo {title} {{Density perturbations and black hole
  formation in hybrid inflation}},\ }\href
  {https://doi.org/10.1103/PhysRevD.54.6040} {\bibfield  {journal} {\bibinfo
  {journal} {Phys. Rev. D}\ }\textbf {\bibinfo {volume} {54}},\ \bibinfo
  {pages} {6040} (\bibinfo {year} {1996})},\ \Eprint
  {https://arxiv.org/abs/astro-ph/9605094} {arXiv:astro-ph/9605094}
  \BibitemShut {NoStop}%
\bibitem [{\citenamefont {Ivanov}(1998)}]{Ivanov:1997ia}%
  \BibitemOpen
  \bibfield  {author} {\bibinfo {author} {\bibfnamefont {P.}~\bibnamefont
  {Ivanov}},\ }\bibfield  {title} {\bibinfo {title} {{Nonlinear metric
  perturbations and production of primordial black holes}},\ }\href
  {https://doi.org/10.1103/PhysRevD.57.7145} {\bibfield  {journal} {\bibinfo
  {journal} {Phys. Rev. D}\ }\textbf {\bibinfo {volume} {57}},\ \bibinfo
  {pages} {7145} (\bibinfo {year} {1998})},\ \Eprint
  {https://arxiv.org/abs/astro-ph/9708224} {arXiv:astro-ph/9708224}
  \BibitemShut {NoStop}%
\bibitem [{\citenamefont {Polnarev}\ and\ \citenamefont
  {Khlopov}(1985)}]{Polnarev:1985btg}%
  \BibitemOpen
  \bibfield  {author} {\bibinfo {author} {\bibfnamefont {A.~G.}\ \bibnamefont
  {Polnarev}}\ and\ \bibinfo {author} {\bibfnamefont {M.~Y.}\ \bibnamefont
  {Khlopov}},\ }\bibfield  {title} {\bibinfo {title} {{COSMOLOGY, PRIMORDIAL
  BLACK HOLES, AND SUPERMASSIVE PARTICLES}},\ }\href
  {https://doi.org/10.1070/PU1985v028n03ABEH003858} {\bibfield  {journal}
  {\bibinfo  {journal} {Sov. Phys. Usp.}\ }\textbf {\bibinfo {volume} {28}},\
  \bibinfo {pages} {213} (\bibinfo {year} {1985})}\BibitemShut {NoStop}%
\bibitem [{\citenamefont {Khlopov}(2010)}]{Khlopov:2008qy}%
  \BibitemOpen
  \bibfield  {author} {\bibinfo {author} {\bibfnamefont {M.~Y.}\ \bibnamefont
  {Khlopov}},\ }\bibfield  {title} {\bibinfo {title} {{Primordial Black
  Holes}},\ }\href {https://doi.org/10.1088/1674-4527/10/6/001} {\bibfield
  {journal} {\bibinfo  {journal} {Res. Astron. Astrophys.}\ }\textbf {\bibinfo
  {volume} {10}},\ \bibinfo {pages} {495} (\bibinfo {year} {2010})},\ \Eprint
  {https://arxiv.org/abs/0801.0116} {arXiv:0801.0116 [astro-ph]} \BibitemShut
  {NoStop}%
\bibitem [{\citenamefont {Sasaki}\ \emph {et~al.}(2016)\citenamefont {Sasaki},
  \citenamefont {Suyama}, \citenamefont {Tanaka},\ and\ \citenamefont
  {Yokoyama}}]{Sasaki:2016jop}%
  \BibitemOpen
  \bibfield  {author} {\bibinfo {author} {\bibfnamefont {M.}~\bibnamefont
  {Sasaki}}, \bibinfo {author} {\bibfnamefont {T.}~\bibnamefont {Suyama}},
  \bibinfo {author} {\bibfnamefont {T.}~\bibnamefont {Tanaka}},\ and\ \bibinfo
  {author} {\bibfnamefont {S.}~\bibnamefont {Yokoyama}},\ }\bibfield  {title}
  {\bibinfo {title} {{Primordial Black Hole Scenario for the Gravitational-Wave
  Event GW150914}},\ }\href {https://doi.org/10.1103/PhysRevLett.117.061101}
  {\bibfield  {journal} {\bibinfo  {journal} {Phys. Rev. Lett.}\ }\textbf
  {\bibinfo {volume} {117}},\ \bibinfo {pages} {061101} (\bibinfo {year}
  {2016})},\ \bibinfo {note} {[Erratum: Phys.Rev.Lett. 121, 059901 (2018)]},\
  \Eprint {https://arxiv.org/abs/1603.08338} {arXiv:1603.08338 [astro-ph.CO]}
  \BibitemShut {NoStop}%
\bibitem [{\citenamefont {Green}\ and\ \citenamefont
  {Kavanagh}(2021)}]{Green:2020jor}%
  \BibitemOpen
  \bibfield  {author} {\bibinfo {author} {\bibfnamefont {A.~M.}\ \bibnamefont
  {Green}}\ and\ \bibinfo {author} {\bibfnamefont {B.~J.}\ \bibnamefont
  {Kavanagh}},\ }\bibfield  {title} {\bibinfo {title} {{Primordial Black Holes
  as a dark matter candidate}},\ }\href
  {https://doi.org/10.1088/1361-6471/abc534} {\bibfield  {journal} {\bibinfo
  {journal} {J. Phys. G}\ }\textbf {\bibinfo {volume} {48}},\ \bibinfo {pages}
  {043001} (\bibinfo {year} {2021})},\ \Eprint
  {https://arxiv.org/abs/2007.10722} {arXiv:2007.10722 [astro-ph.CO]}
  \BibitemShut {NoStop}%
\bibitem [{\citenamefont {Carr}\ \emph {et~al.}(2017)\citenamefont {Carr},
  \citenamefont {Raidal}, \citenamefont {Tenkanen}, \citenamefont {Vaskonen},\
  and\ \citenamefont {Veerm\"ae}}]{Carr:2017jsz}%
  \BibitemOpen
  \bibfield  {author} {\bibinfo {author} {\bibfnamefont {B.}~\bibnamefont
  {Carr}}, \bibinfo {author} {\bibfnamefont {M.}~\bibnamefont {Raidal}},
  \bibinfo {author} {\bibfnamefont {T.}~\bibnamefont {Tenkanen}}, \bibinfo
  {author} {\bibfnamefont {V.}~\bibnamefont {Vaskonen}},\ and\ \bibinfo
  {author} {\bibfnamefont {H.}~\bibnamefont {Veerm\"ae}},\ }\bibfield  {title}
  {\bibinfo {title} {{Primordial black hole constraints for extended mass
  functions}},\ }\href {https://doi.org/10.1103/PhysRevD.96.023514} {\bibfield
  {journal} {\bibinfo  {journal} {Phys. Rev. D}\ }\textbf {\bibinfo {volume}
  {96}},\ \bibinfo {pages} {023514} (\bibinfo {year} {2017})},\ \Eprint
  {https://arxiv.org/abs/1705.05567} {arXiv:1705.05567 [astro-ph.CO]}
  \BibitemShut {NoStop}%
\bibitem [{\citenamefont {Ali-Ha\"\i{}moud}\ \emph {et~al.}(2017)\citenamefont
  {Ali-Ha\"\i{}moud}, \citenamefont {Kovetz},\ and\ \citenamefont
  {Kamionkowski}}]{AliHaimoud:2017rtz}%
  \BibitemOpen
  \bibfield  {author} {\bibinfo {author} {\bibfnamefont {Y.}~\bibnamefont
  {Ali-Ha\"\i{}moud}}, \bibinfo {author} {\bibfnamefont {E.~D.}\ \bibnamefont
  {Kovetz}},\ and\ \bibinfo {author} {\bibfnamefont {M.}~\bibnamefont
  {Kamionkowski}},\ }\bibfield  {title} {\bibinfo {title} {{Merger rate of
  primordial black-hole binaries}},\ }\href
  {https://doi.org/10.1103/PhysRevD.96.123523} {\bibfield  {journal} {\bibinfo
  {journal} {Phys. Rev. D}\ }\textbf {\bibinfo {volume} {96}},\ \bibinfo
  {pages} {123523} (\bibinfo {year} {2017})},\ \Eprint
  {https://arxiv.org/abs/1709.06576} {arXiv:1709.06576 [astro-ph.CO]}
  \BibitemShut {NoStop}%
\bibitem [{\citenamefont {Nakamura}\ \emph {et~al.}(1997)\citenamefont
  {Nakamura}, \citenamefont {Sasaki}, \citenamefont {Tanaka},\ and\
  \citenamefont {Thorne}}]{Nakamura:1997sm}%
  \BibitemOpen
  \bibfield  {author} {\bibinfo {author} {\bibfnamefont {T.}~\bibnamefont
  {Nakamura}}, \bibinfo {author} {\bibfnamefont {M.}~\bibnamefont {Sasaki}},
  \bibinfo {author} {\bibfnamefont {T.}~\bibnamefont {Tanaka}},\ and\ \bibinfo
  {author} {\bibfnamefont {K.~S.}\ \bibnamefont {Thorne}},\ }\bibfield  {title}
  {\bibinfo {title} {{Gravitational waves from coalescing black hole MACHO
  binaries}},\ }\href {https://doi.org/10.1086/310886} {\bibfield  {journal}
  {\bibinfo  {journal} {Astrophys. J. Lett.}\ }\textbf {\bibinfo {volume}
  {487}},\ \bibinfo {pages} {L139} (\bibinfo {year} {1997})},\ \Eprint
  {https://arxiv.org/abs/astro-ph/9708060} {arXiv:astro-ph/9708060}
  \BibitemShut {NoStop}%
\bibitem [{\citenamefont {Ioka}\ \emph {et~al.}(1998)\citenamefont {Ioka},
  \citenamefont {Chiba}, \citenamefont {Tanaka},\ and\ \citenamefont
  {Nakamura}}]{Ioka:1998nz}%
  \BibitemOpen
  \bibfield  {author} {\bibinfo {author} {\bibfnamefont {K.}~\bibnamefont
  {Ioka}}, \bibinfo {author} {\bibfnamefont {T.}~\bibnamefont {Chiba}},
  \bibinfo {author} {\bibfnamefont {T.}~\bibnamefont {Tanaka}},\ and\ \bibinfo
  {author} {\bibfnamefont {T.}~\bibnamefont {Nakamura}},\ }\bibfield  {title}
  {\bibinfo {title} {{Black hole binary formation in the expanding universe:
  Three body problem approximation}},\ }\href
  {https://doi.org/10.1103/PhysRevD.58.063003} {\bibfield  {journal} {\bibinfo
  {journal} {Phys. Rev. D}\ }\textbf {\bibinfo {volume} {58}},\ \bibinfo
  {pages} {063003} (\bibinfo {year} {1998})},\ \Eprint
  {https://arxiv.org/abs/astro-ph/9807018} {arXiv:astro-ph/9807018}
  \BibitemShut {NoStop}%
\bibitem [{\citenamefont {Aasi}\ \emph {et~al.}(2015)\citenamefont {Aasi} \emph
  {et~al.}}]{TheLIGOScientific:2014jea}%
  \BibitemOpen
  \bibfield  {author} {\bibinfo {author} {\bibfnamefont {J.}~\bibnamefont
  {Aasi}} \emph {et~al.} (\bibinfo {collaboration} {LIGO Scientific}),\
  }\bibfield  {title} {\bibinfo {title} {{Advanced LIGO}},\ }\href
  {https://doi.org/10.1088/0264-9381/32/7/074001} {\bibfield  {journal}
  {\bibinfo  {journal} {Class. Quant. Grav.}\ }\textbf {\bibinfo {volume}
  {32}},\ \bibinfo {pages} {074001} (\bibinfo {year} {2015})},\ \Eprint
  {https://arxiv.org/abs/1411.4547} {arXiv:1411.4547 [gr-qc]} \BibitemShut
  {NoStop}%
\bibitem [{\citenamefont {Acernese}\ \emph {et~al.}(2015)\citenamefont
  {Acernese} \emph {et~al.}}]{TheVirgo:2014hva}%
  \BibitemOpen
  \bibfield  {author} {\bibinfo {author} {\bibfnamefont {F.}~\bibnamefont
  {Acernese}} \emph {et~al.} (\bibinfo {collaboration} {VIRGO}),\ }\bibfield
  {title} {\bibinfo {title} {{Advanced Virgo: a second-generation
  interferometric gravitational wave detector}},\ }\href
  {https://doi.org/10.1088/0264-9381/32/2/024001} {\bibfield  {journal}
  {\bibinfo  {journal} {Class. Quant. Grav.}\ }\textbf {\bibinfo {volume}
  {32}},\ \bibinfo {pages} {024001} (\bibinfo {year} {2015})},\ \Eprint
  {https://arxiv.org/abs/1408.3978} {arXiv:1408.3978 [gr-qc]} \BibitemShut
  {NoStop}%
\bibitem [{\citenamefont {Aso}\ \emph {et~al.}(2013)\citenamefont {Aso},
  \citenamefont {Michimura}, \citenamefont {Somiya}, \citenamefont {Ando},
  \citenamefont {Miyakawa}, \citenamefont {Sekiguchi}, \citenamefont
  {Tatsumi},\ and\ \citenamefont {Yamamoto}}]{Aso:2013eba}%
  \BibitemOpen
  \bibfield  {author} {\bibinfo {author} {\bibfnamefont {Y.}~\bibnamefont
  {Aso}}, \bibinfo {author} {\bibfnamefont {Y.}~\bibnamefont {Michimura}},
  \bibinfo {author} {\bibfnamefont {K.}~\bibnamefont {Somiya}}, \bibinfo
  {author} {\bibfnamefont {M.}~\bibnamefont {Ando}}, \bibinfo {author}
  {\bibfnamefont {O.}~\bibnamefont {Miyakawa}}, \bibinfo {author}
  {\bibfnamefont {T.}~\bibnamefont {Sekiguchi}}, \bibinfo {author}
  {\bibfnamefont {D.}~\bibnamefont {Tatsumi}},\ and\ \bibinfo {author}
  {\bibfnamefont {H.}~\bibnamefont {Yamamoto}} (\bibinfo {collaboration}
  {KAGRA}),\ }\bibfield  {title} {\bibinfo {title} {{Interferometer design of
  the KAGRA gravitational wave detector}},\ }\href
  {https://doi.org/10.1103/PhysRevD.88.043007} {\bibfield  {journal} {\bibinfo
  {journal} {Phys. Rev. D}\ }\textbf {\bibinfo {volume} {88}},\ \bibinfo
  {pages} {043007} (\bibinfo {year} {2013})},\ \Eprint
  {https://arxiv.org/abs/1306.6747} {arXiv:1306.6747 [gr-qc]} \BibitemShut
  {NoStop}%
\bibitem [{\citenamefont {Clesse}\ and\ \citenamefont
  {Garc\'\i{}a-Bellido}(2017)}]{Clesse:2016vqa}%
  \BibitemOpen
  \bibfield  {author} {\bibinfo {author} {\bibfnamefont {S.}~\bibnamefont
  {Clesse}}\ and\ \bibinfo {author} {\bibfnamefont {J.}~\bibnamefont
  {Garc\'\i{}a-Bellido}},\ }\bibfield  {title} {\bibinfo {title} {{The
  clustering of massive Primordial Black Holes as Dark Matter: measuring their
  mass distribution with Advanced LIGO}},\ }\href
  {https://doi.org/10.1016/j.dark.2016.10.002} {\bibfield  {journal} {\bibinfo
  {journal} {Phys. Dark Univ.}\ }\textbf {\bibinfo {volume} {15}},\ \bibinfo
  {pages} {142} (\bibinfo {year} {2017})},\ \Eprint
  {https://arxiv.org/abs/1603.05234} {arXiv:1603.05234 [astro-ph.CO]}
  \BibitemShut {NoStop}%
\bibitem [{\citenamefont {Raidal}\ \emph {et~al.}(2017)\citenamefont {Raidal},
  \citenamefont {Vaskonen},\ and\ \citenamefont {Veerm\"ae}}]{Raidal:2017mfl}%
  \BibitemOpen
  \bibfield  {author} {\bibinfo {author} {\bibfnamefont {M.}~\bibnamefont
  {Raidal}}, \bibinfo {author} {\bibfnamefont {V.}~\bibnamefont {Vaskonen}},\
  and\ \bibinfo {author} {\bibfnamefont {H.}~\bibnamefont {Veerm\"ae}},\
  }\bibfield  {title} {\bibinfo {title} {{Gravitational Waves from Primordial
  Black Hole Mergers}},\ }\href {https://doi.org/10.1088/1475-7516/2017/09/037}
  {\bibfield  {journal} {\bibinfo  {journal} {JCAP}\ }\textbf {\bibinfo
  {volume} {09}},\ \bibinfo {pages} {037}},\ \Eprint
  {https://arxiv.org/abs/1707.01480} {arXiv:1707.01480 [astro-ph.CO]}
  \BibitemShut {NoStop}%
\bibitem [{\citenamefont {Belotsky}\ \emph {et~al.}(2019)\citenamefont
  {Belotsky}, \citenamefont {Dokuchaev}, \citenamefont {Eroshenko},
  \citenamefont {Esipova}, \citenamefont {Khlopov}, \citenamefont {Khromykh},
  \citenamefont {Kirillov}, \citenamefont {Nikulin}, \citenamefont {Rubin},\
  and\ \citenamefont {Svadkovsky}}]{Belotsky:2018wph}%
  \BibitemOpen
  \bibfield  {author} {\bibinfo {author} {\bibfnamefont {K.~M.}\ \bibnamefont
  {Belotsky}}, \bibinfo {author} {\bibfnamefont {V.~I.}\ \bibnamefont
  {Dokuchaev}}, \bibinfo {author} {\bibfnamefont {Y.~N.}\ \bibnamefont
  {Eroshenko}}, \bibinfo {author} {\bibfnamefont {E.~A.}\ \bibnamefont
  {Esipova}}, \bibinfo {author} {\bibfnamefont {M.~Y.}\ \bibnamefont
  {Khlopov}}, \bibinfo {author} {\bibfnamefont {L.~A.}\ \bibnamefont
  {Khromykh}}, \bibinfo {author} {\bibfnamefont {A.~A.}\ \bibnamefont
  {Kirillov}}, \bibinfo {author} {\bibfnamefont {V.~V.}\ \bibnamefont
  {Nikulin}}, \bibinfo {author} {\bibfnamefont {S.~G.}\ \bibnamefont {Rubin}},\
  and\ \bibinfo {author} {\bibfnamefont {I.~V.}\ \bibnamefont {Svadkovsky}},\
  }\bibfield  {title} {\bibinfo {title} {{Clusters of primordial black
  holes}},\ }\href {https://doi.org/10.1140/epjc/s10052-019-6741-4} {\bibfield
  {journal} {\bibinfo  {journal} {Eur. Phys. J. C}\ }\textbf {\bibinfo {volume}
  {79}},\ \bibinfo {pages} {246} (\bibinfo {year} {2019})},\ \Eprint
  {https://arxiv.org/abs/1807.06590} {arXiv:1807.06590 [astro-ph.CO]}
  \BibitemShut {NoStop}%
\bibitem [{\citenamefont {Chen}\ and\ \citenamefont
  {Huang}(2018)}]{Chen:2018czv}%
  \BibitemOpen
  \bibfield  {author} {\bibinfo {author} {\bibfnamefont {Z.-C.}\ \bibnamefont
  {Chen}}\ and\ \bibinfo {author} {\bibfnamefont {Q.-G.}\ \bibnamefont
  {Huang}},\ }\bibfield  {title} {\bibinfo {title} {{Merger Rate Distribution
  of Primordial-Black-Hole Binaries}},\ }\href
  {https://doi.org/10.3847/1538-4357/aad6e2} {\bibfield  {journal} {\bibinfo
  {journal} {Astrophys. J.}\ }\textbf {\bibinfo {volume} {864}},\ \bibinfo
  {pages} {61} (\bibinfo {year} {2018})},\ \Eprint
  {https://arxiv.org/abs/1801.10327} {arXiv:1801.10327 [astro-ph.CO]}
  \BibitemShut {NoStop}%
\bibitem [{\citenamefont {Raidal}\ \emph {et~al.}(2019)\citenamefont {Raidal},
  \citenamefont {Spethmann}, \citenamefont {Vaskonen},\ and\ \citenamefont
  {Veerm\"ae}}]{Raidal:2018bbj}%
  \BibitemOpen
  \bibfield  {author} {\bibinfo {author} {\bibfnamefont {M.}~\bibnamefont
  {Raidal}}, \bibinfo {author} {\bibfnamefont {C.}~\bibnamefont {Spethmann}},
  \bibinfo {author} {\bibfnamefont {V.}~\bibnamefont {Vaskonen}},\ and\
  \bibinfo {author} {\bibfnamefont {H.}~\bibnamefont {Veerm\"ae}},\ }\bibfield
  {title} {\bibinfo {title} {{Formation and Evolution of Primordial Black Hole
  Binaries in the Early Universe}},\ }\href
  {https://doi.org/10.1088/1475-7516/2019/02/018} {\bibfield  {journal}
  {\bibinfo  {journal} {JCAP}\ }\textbf {\bibinfo {volume} {02}},\ \bibinfo
  {pages} {018}},\ \Eprint {https://arxiv.org/abs/1812.01930} {arXiv:1812.01930
  [astro-ph.CO]} \BibitemShut {NoStop}%
\bibitem [{\citenamefont {De~Luca}\ \emph
  {et~al.}(2020{\natexlab{a}})\citenamefont {De~Luca}, \citenamefont
  {Franciolini}, \citenamefont {Pani},\ and\ \citenamefont
  {Riotto}}]{DeLuca:2020fpg}%
  \BibitemOpen
  \bibfield  {author} {\bibinfo {author} {\bibfnamefont {V.}~\bibnamefont
  {De~Luca}}, \bibinfo {author} {\bibfnamefont {G.}~\bibnamefont
  {Franciolini}}, \bibinfo {author} {\bibfnamefont {P.}~\bibnamefont {Pani}},\
  and\ \bibinfo {author} {\bibfnamefont {A.}~\bibnamefont {Riotto}},\
  }\bibfield  {title} {\bibinfo {title} {{Constraints on Primordial Black
  Holes: the Importance of Accretion}},\ }\href
  {https://doi.org/10.1103/PhysRevD.102.043505} {\bibfield  {journal} {\bibinfo
   {journal} {Phys. Rev. D}\ }\textbf {\bibinfo {volume} {102}},\ \bibinfo
  {pages} {043505} (\bibinfo {year} {2020}{\natexlab{a}})},\ \Eprint
  {https://arxiv.org/abs/2003.12589} {arXiv:2003.12589 [astro-ph.CO]}
  \BibitemShut {NoStop}%
\bibitem [{\citenamefont {De~Luca}\ \emph
  {et~al.}(2020{\natexlab{b}})\citenamefont {De~Luca}, \citenamefont
  {Franciolini}, \citenamefont {Pani},\ and\ \citenamefont
  {Riotto}}]{DeLuca:2020bjf}%
  \BibitemOpen
  \bibfield  {author} {\bibinfo {author} {\bibfnamefont {V.}~\bibnamefont
  {De~Luca}}, \bibinfo {author} {\bibfnamefont {G.}~\bibnamefont
  {Franciolini}}, \bibinfo {author} {\bibfnamefont {P.}~\bibnamefont {Pani}},\
  and\ \bibinfo {author} {\bibfnamefont {A.}~\bibnamefont {Riotto}},\
  }\bibfield  {title} {\bibinfo {title} {{The evolution of primordial black
  holes and their final observable spins}},\ }\href
  {https://doi.org/10.1088/1475-7516/2020/04/052} {\bibfield  {journal}
  {\bibinfo  {journal} {JCAP}\ }\textbf {\bibinfo {volume} {04}},\ \bibinfo
  {pages} {052}},\ \Eprint {https://arxiv.org/abs/2003.02778} {arXiv:2003.02778
  [astro-ph.CO]} \BibitemShut {NoStop}%
\bibitem [{\citenamefont {Abbott}\ \emph {et~al.}(2020)\citenamefont {Abbott}
  \emph {et~al.}}]{GWTC2}%
  \BibitemOpen
  \bibfield  {author} {\bibinfo {author} {\bibfnamefont {R.}~\bibnamefont
  {Abbott}} \emph {et~al.} (\bibinfo {collaboration} {LIGO Scientific,
  Virgo}),\ }\bibfield  {title} {\bibinfo {title} {{GWTC-2: Compact Binary
  Coalescences Observed by LIGO and Virgo During the First Half of the Third
  Observing Run}},\ }\href@noop {} {\bibfield  {journal} {\bibinfo  {journal}
  {arXiv e-print}\ } (\bibinfo {year} {2020})},\ \Eprint
  {https://arxiv.org/abs/2010.14527} {arXiv:2010.14527 [gr-qc]} \BibitemShut
  {NoStop}%
\bibitem [{\citenamefont {Wong}\ \emph {et~al.}(2021)\citenamefont {Wong},
  \citenamefont {Franciolini}, \citenamefont {De~Luca}, \citenamefont
  {Baibhav}, \citenamefont {Berti}, \citenamefont {Pani},\ and\ \citenamefont
  {Riotto}}]{Wong:2020yig}%
  \BibitemOpen
  \bibfield  {author} {\bibinfo {author} {\bibfnamefont {K.~W.~K.}\
  \bibnamefont {Wong}}, \bibinfo {author} {\bibfnamefont {G.}~\bibnamefont
  {Franciolini}}, \bibinfo {author} {\bibfnamefont {V.}~\bibnamefont
  {De~Luca}}, \bibinfo {author} {\bibfnamefont {V.}~\bibnamefont {Baibhav}},
  \bibinfo {author} {\bibfnamefont {E.}~\bibnamefont {Berti}}, \bibinfo
  {author} {\bibfnamefont {P.}~\bibnamefont {Pani}},\ and\ \bibinfo {author}
  {\bibfnamefont {A.}~\bibnamefont {Riotto}},\ }\bibfield  {title} {\bibinfo
  {title} {{Constraining the primordial black hole scenario with Bayesian
  inference and machine learning: the GWTC-2 gravitational wave catalog}},\
  }\href {https://doi.org/10.1103/PhysRevD.103.023026} {\bibfield  {journal}
  {\bibinfo  {journal} {Phys. Rev. D}\ }\textbf {\bibinfo {volume} {103}},\
  \bibinfo {pages} {023026} (\bibinfo {year} {2021})},\ \Eprint
  {https://arxiv.org/abs/2011.01865} {arXiv:2011.01865 [gr-qc]} \BibitemShut
  {NoStop}%
\bibitem [{\citenamefont {De~Luca}\ \emph
  {et~al.}(2020{\natexlab{c}})\citenamefont {De~Luca}, \citenamefont
  {Franciolini}, \citenamefont {Pani},\ and\ \citenamefont
  {Riotto}}]{DeLuca:2020qqa}%
  \BibitemOpen
  \bibfield  {author} {\bibinfo {author} {\bibfnamefont {V.}~\bibnamefont
  {De~Luca}}, \bibinfo {author} {\bibfnamefont {G.}~\bibnamefont
  {Franciolini}}, \bibinfo {author} {\bibfnamefont {P.}~\bibnamefont {Pani}},\
  and\ \bibinfo {author} {\bibfnamefont {A.}~\bibnamefont {Riotto}},\
  }\bibfield  {title} {\bibinfo {title} {{Primordial Black Holes Confront
  LIGO/Virgo data: Current situation}},\ }\href
  {https://doi.org/10.1088/1475-7516/2020/06/044} {\bibfield  {journal}
  {\bibinfo  {journal} {JCAP}\ }\textbf {\bibinfo {volume} {06}},\ \bibinfo
  {pages} {044}},\ \Eprint {https://arxiv.org/abs/2005.05641} {arXiv:2005.05641
  [astro-ph.CO]} \BibitemShut {NoStop}%
\bibitem [{\citenamefont {H\"utsi}\ \emph {et~al.}(2021)\citenamefont
  {H\"utsi}, \citenamefont {Raidal}, \citenamefont {Vaskonen},\ and\
  \citenamefont {Veerm\"ae}}]{Hutsi:2020sol}%
  \BibitemOpen
  \bibfield  {author} {\bibinfo {author} {\bibfnamefont {G.}~\bibnamefont
  {H\"utsi}}, \bibinfo {author} {\bibfnamefont {M.}~\bibnamefont {Raidal}},
  \bibinfo {author} {\bibfnamefont {V.}~\bibnamefont {Vaskonen}},\ and\
  \bibinfo {author} {\bibfnamefont {H.}~\bibnamefont {Veerm\"ae}},\ }\bibfield
  {title} {\bibinfo {title} {{Two populations of LIGO-Virgo black holes}},\
  }\href {https://doi.org/10.1088/1475-7516/2021/03/068} {\bibfield  {journal}
  {\bibinfo  {journal} {JCAP}\ }\textbf {\bibinfo {volume} {03}},\ \bibinfo
  {pages} {068}},\ \Eprint {https://arxiv.org/abs/2012.02786} {arXiv:2012.02786
  [astro-ph.CO]} \BibitemShut {NoStop}%
\bibitem [{\citenamefont {Franciolini}\ \emph {et~al.}(2021)\citenamefont
  {Franciolini}, \citenamefont {Baibhav}, \citenamefont {De~Luca},
  \citenamefont {Ng}, \citenamefont {Wong}, \citenamefont {Berti},
  \citenamefont {Pani}, \citenamefont {Riotto},\ and\ \citenamefont
  {Vitale}}]{Franciolini:2021tla}%
  \BibitemOpen
  \bibfield  {author} {\bibinfo {author} {\bibfnamefont {G.}~\bibnamefont
  {Franciolini}}, \bibinfo {author} {\bibfnamefont {V.}~\bibnamefont
  {Baibhav}}, \bibinfo {author} {\bibfnamefont {V.}~\bibnamefont {De~Luca}},
  \bibinfo {author} {\bibfnamefont {K.~K.~Y.}\ \bibnamefont {Ng}}, \bibinfo
  {author} {\bibfnamefont {K.~W.~K.}\ \bibnamefont {Wong}}, \bibinfo {author}
  {\bibfnamefont {E.}~\bibnamefont {Berti}}, \bibinfo {author} {\bibfnamefont
  {P.}~\bibnamefont {Pani}}, \bibinfo {author} {\bibfnamefont {A.}~\bibnamefont
  {Riotto}},\ and\ \bibinfo {author} {\bibfnamefont {S.}~\bibnamefont
  {Vitale}},\ }\bibfield  {title} {\bibinfo {title} {{Quantifying the evidence
  for primordial black holes in LIGO/Virgo gravitational-wave data}},\
  }\href@noop {} {\bibfield  {journal} {\bibinfo  {journal} {arXiv}\ }
  (\bibinfo {year} {2021})},\ \Eprint {https://arxiv.org/abs/2105.03349}
  {arXiv:2105.03349 [gr-qc]} \BibitemShut {NoStop}%
\bibitem [{\citenamefont {Mukherjee}\ and\ \citenamefont
  {Silk}(2021)}]{Mukherjee:2021ags}%
  \BibitemOpen
  \bibfield  {author} {\bibinfo {author} {\bibfnamefont {S.}~\bibnamefont
  {Mukherjee}}\ and\ \bibinfo {author} {\bibfnamefont {J.}~\bibnamefont
  {Silk}},\ }\bibfield  {title} {\bibinfo {title} {{Can we distinguish
  astrophysical from primordial black holes via the stochastic gravitational
  wave background?}},\ }\href {https://doi.org/10.1093/mnras/stab1932}
  {\bibfield  {journal} {\bibinfo  {journal} {Mon. Not. Roy. Astron. Soc.}\
  }\textbf {\bibinfo {volume} {506}},\ \bibinfo {pages} {3977} (\bibinfo {year}
  {2021})},\ \Eprint {https://arxiv.org/abs/2105.11139} {arXiv:2105.11139
  [gr-qc]} \BibitemShut {NoStop}%
\bibitem [{\citenamefont {Hall}\ and\ \citenamefont
  {Evans}(2019)}]{Hall:2019xmm}%
  \BibitemOpen
  \bibfield  {author} {\bibinfo {author} {\bibfnamefont {E.~D.}\ \bibnamefont
  {Hall}}\ and\ \bibinfo {author} {\bibfnamefont {M.}~\bibnamefont {Evans}},\
  }\bibfield  {title} {\bibinfo {title} {{Metrics for next-generation
  gravitational-wave detectors}},\ }\href
  {https://doi.org/10.1088/1361-6382/ab41d6} {\bibfield  {journal} {\bibinfo
  {journal} {Class. Quant. Grav.}\ }\textbf {\bibinfo {volume} {36}},\ \bibinfo
  {pages} {225002} (\bibinfo {year} {2019})},\ \Eprint
  {https://arxiv.org/abs/1902.09485} {arXiv:1902.09485 [astro-ph.IM]}
  \BibitemShut {NoStop}%
\bibitem [{\citenamefont {O'Shaughnessy}\ \emph {et~al.}(2017)\citenamefont
  {O'Shaughnessy}, \citenamefont {Bellovary}, \citenamefont {Brooks},
  \citenamefont {Shen}, \citenamefont {Governato},\ and\ \citenamefont
  {Christensen}}]{OShaughnessy:2016nny}%
  \BibitemOpen
  \bibfield  {author} {\bibinfo {author} {\bibfnamefont {R.}~\bibnamefont
  {O'Shaughnessy}}, \bibinfo {author} {\bibfnamefont {J.}~\bibnamefont
  {Bellovary}}, \bibinfo {author} {\bibfnamefont {A.}~\bibnamefont {Brooks}},
  \bibinfo {author} {\bibfnamefont {S.}~\bibnamefont {Shen}}, \bibinfo {author}
  {\bibfnamefont {F.}~\bibnamefont {Governato}},\ and\ \bibinfo {author}
  {\bibfnamefont {C.}~\bibnamefont {Christensen}},\ }\bibfield  {title}
  {\bibinfo {title} {{The effects of host galaxy properties on merging compact
  binaries detectable by LIGO}},\ }\href
  {https://doi.org/10.1093/mnras/stw2550} {\bibfield  {journal} {\bibinfo
  {journal} {Mon. Not. Roy. Astron. Soc.}\ }\textbf {\bibinfo {volume} {464}},\
  \bibinfo {pages} {2831} (\bibinfo {year} {2017})},\ \Eprint
  {https://arxiv.org/abs/1609.06715} {arXiv:1609.06715 [astro-ph.GA]}
  \BibitemShut {NoStop}%
\bibitem [{\citenamefont {Dominik}\ \emph {et~al.}(2012)\citenamefont
  {Dominik}, \citenamefont {Belczynski}, \citenamefont {Fryer}, \citenamefont
  {Holz}, \citenamefont {Berti}, \citenamefont {Bulik}, \citenamefont
  {Mandel},\ and\ \citenamefont {O'Shaughnessy}}]{Dominik:2012kk}%
  \BibitemOpen
  \bibfield  {author} {\bibinfo {author} {\bibfnamefont {M.}~\bibnamefont
  {Dominik}}, \bibinfo {author} {\bibfnamefont {K.}~\bibnamefont {Belczynski}},
  \bibinfo {author} {\bibfnamefont {C.}~\bibnamefont {Fryer}}, \bibinfo
  {author} {\bibfnamefont {D.}~\bibnamefont {Holz}}, \bibinfo {author}
  {\bibfnamefont {E.}~\bibnamefont {Berti}}, \bibinfo {author} {\bibfnamefont
  {T.}~\bibnamefont {Bulik}}, \bibinfo {author} {\bibfnamefont
  {I.}~\bibnamefont {Mandel}},\ and\ \bibinfo {author} {\bibfnamefont
  {R.}~\bibnamefont {O'Shaughnessy}},\ }\bibfield  {title} {\bibinfo {title}
  {{Double Compact Objects I: The Significance of the Common Envelope on Merger
  Rates}},\ }\href {https://doi.org/10.1088/0004-637X/759/1/52} {\bibfield
  {journal} {\bibinfo  {journal} {Astrophys. J.}\ }\textbf {\bibinfo {volume}
  {759}},\ \bibinfo {pages} {52} (\bibinfo {year} {2012})},\ \Eprint
  {https://arxiv.org/abs/1202.4901} {arXiv:1202.4901 [astro-ph.HE]}
  \BibitemShut {NoStop}%
\bibitem [{\citenamefont {Dominik}\ \emph {et~al.}(2013)\citenamefont
  {Dominik}, \citenamefont {Belczynski}, \citenamefont {Fryer}, \citenamefont
  {Holz}, \citenamefont {Berti}, \citenamefont {Bulik}, \citenamefont
  {Mandel},\ and\ \citenamefont {O'Shaughnessy}}]{Dominik:2013tma}%
  \BibitemOpen
  \bibfield  {author} {\bibinfo {author} {\bibfnamefont {M.}~\bibnamefont
  {Dominik}}, \bibinfo {author} {\bibfnamefont {K.}~\bibnamefont {Belczynski}},
  \bibinfo {author} {\bibfnamefont {C.}~\bibnamefont {Fryer}}, \bibinfo
  {author} {\bibfnamefont {D.~E.}\ \bibnamefont {Holz}}, \bibinfo {author}
  {\bibfnamefont {E.}~\bibnamefont {Berti}}, \bibinfo {author} {\bibfnamefont
  {T.}~\bibnamefont {Bulik}}, \bibinfo {author} {\bibfnamefont
  {I.}~\bibnamefont {Mandel}},\ and\ \bibinfo {author} {\bibfnamefont
  {R.}~\bibnamefont {O'Shaughnessy}},\ }\bibfield  {title} {\bibinfo {title}
  {{Double Compact Objects II: Cosmological Merger Rates}},\ }\href
  {https://doi.org/10.1088/0004-637X/779/1/72} {\bibfield  {journal} {\bibinfo
  {journal} {Astrophys. J.}\ }\textbf {\bibinfo {volume} {779}},\ \bibinfo
  {pages} {72} (\bibinfo {year} {2013})},\ \Eprint
  {https://arxiv.org/abs/1308.1546} {arXiv:1308.1546 [astro-ph.HE]}
  \BibitemShut {NoStop}%
\bibitem [{\citenamefont {Dominik}\ \emph {et~al.}(2015)\citenamefont
  {Dominik}, \citenamefont {Berti}, \citenamefont {O'Shaughnessy},
  \citenamefont {Mandel}, \citenamefont {Belczynski}, \citenamefont {Fryer},
  \citenamefont {Holz}, \citenamefont {Bulik},\ and\ \citenamefont
  {Pannarale}}]{Dominik:2014yma}%
  \BibitemOpen
  \bibfield  {author} {\bibinfo {author} {\bibfnamefont {M.}~\bibnamefont
  {Dominik}}, \bibinfo {author} {\bibfnamefont {E.}~\bibnamefont {Berti}},
  \bibinfo {author} {\bibfnamefont {R.}~\bibnamefont {O'Shaughnessy}}, \bibinfo
  {author} {\bibfnamefont {I.}~\bibnamefont {Mandel}}, \bibinfo {author}
  {\bibfnamefont {K.}~\bibnamefont {Belczynski}}, \bibinfo {author}
  {\bibfnamefont {C.}~\bibnamefont {Fryer}}, \bibinfo {author} {\bibfnamefont
  {D.~E.}\ \bibnamefont {Holz}}, \bibinfo {author} {\bibfnamefont
  {T.}~\bibnamefont {Bulik}},\ and\ \bibinfo {author} {\bibfnamefont
  {F.}~\bibnamefont {Pannarale}},\ }\bibfield  {title} {\bibinfo {title}
  {{Double Compact Objects III: Gravitational Wave Detection Rates}},\ }\href
  {https://doi.org/10.1088/0004-637X/806/2/263} {\bibfield  {journal} {\bibinfo
   {journal} {Astrophys. J.}\ }\textbf {\bibinfo {volume} {806}},\ \bibinfo
  {pages} {263} (\bibinfo {year} {2015})},\ \Eprint
  {https://arxiv.org/abs/1405.7016} {arXiv:1405.7016 [astro-ph.HE]}
  \BibitemShut {NoStop}%
\bibitem [{\citenamefont {de~Mink}\ and\ \citenamefont
  {Belczynski}(2015)}]{deMink:2015yea}%
  \BibitemOpen
  \bibfield  {author} {\bibinfo {author} {\bibfnamefont {S.}~\bibnamefont
  {de~Mink}}\ and\ \bibinfo {author} {\bibfnamefont {K.}~\bibnamefont
  {Belczynski}},\ }\bibfield  {title} {\bibinfo {title} {{Merger rates of
  double neutron stars and stellar origin black holes: The Impact of Initial
  Conditions on Binary Evolution Predictions}},\ }\href
  {https://doi.org/10.1088/0004-637X/814/1/58} {\bibfield  {journal} {\bibinfo
  {journal} {Astrophys. J.}\ }\textbf {\bibinfo {volume} {814}},\ \bibinfo
  {pages} {58} (\bibinfo {year} {2015})},\ \Eprint
  {https://arxiv.org/abs/1506.03573} {arXiv:1506.03573 [astro-ph.HE]}
  \BibitemShut {NoStop}%
\bibitem [{\citenamefont {Belczynski}\ \emph {et~al.}(2016)\citenamefont
  {Belczynski}, \citenamefont {Holz}, \citenamefont {Bulik},\ and\
  \citenamefont {O'Shaughnessy}}]{Belczynski:2016obo}%
  \BibitemOpen
  \bibfield  {author} {\bibinfo {author} {\bibfnamefont {K.}~\bibnamefont
  {Belczynski}}, \bibinfo {author} {\bibfnamefont {D.~E.}\ \bibnamefont
  {Holz}}, \bibinfo {author} {\bibfnamefont {T.}~\bibnamefont {Bulik}},\ and\
  \bibinfo {author} {\bibfnamefont {R.}~\bibnamefont {O'Shaughnessy}},\
  }\bibfield  {title} {\bibinfo {title} {{The first gravitational-wave source
  from the isolated evolution of two 40-100 Msun stars}},\ }\href
  {https://doi.org/10.1038/nature18322} {\bibfield  {journal} {\bibinfo
  {journal} {Nature}\ }\textbf {\bibinfo {volume} {534}},\ \bibinfo {pages}
  {512} (\bibinfo {year} {2016})},\ \Eprint {https://arxiv.org/abs/1602.04531}
  {arXiv:1602.04531 [astro-ph.HE]} \BibitemShut {NoStop}%
\bibitem [{\citenamefont {Stevenson}\ \emph {et~al.}(2017)\citenamefont
  {Stevenson}, \citenamefont {Vigna-G\'omez}, \citenamefont {Mandel},
  \citenamefont {Barrett}, \citenamefont {Neijssel}, \citenamefont {Perkins},\
  and\ \citenamefont {de~Mink}}]{Stevenson:2017tfq}%
  \BibitemOpen
  \bibfield  {author} {\bibinfo {author} {\bibfnamefont {S.}~\bibnamefont
  {Stevenson}}, \bibinfo {author} {\bibfnamefont {A.}~\bibnamefont
  {Vigna-G\'omez}}, \bibinfo {author} {\bibfnamefont {I.}~\bibnamefont
  {Mandel}}, \bibinfo {author} {\bibfnamefont {J.~W.}\ \bibnamefont {Barrett}},
  \bibinfo {author} {\bibfnamefont {C.~J.}\ \bibnamefont {Neijssel}}, \bibinfo
  {author} {\bibfnamefont {D.}~\bibnamefont {Perkins}},\ and\ \bibinfo {author}
  {\bibfnamefont {S.~E.}\ \bibnamefont {de~Mink}},\ }\bibfield  {title}
  {\bibinfo {title} {{Formation of the first three gravitational-wave
  observations through isolated binary evolution}},\ }\href
  {https://doi.org/10.1038/ncomms14906} {\bibfield  {journal} {\bibinfo
  {journal} {Nature Commun.}\ }\textbf {\bibinfo {volume} {8}},\ \bibinfo
  {pages} {14906} (\bibinfo {year} {2017})},\ \Eprint
  {https://arxiv.org/abs/1704.01352} {arXiv:1704.01352 [astro-ph.HE]}
  \BibitemShut {NoStop}%
\bibitem [{\citenamefont {Mapelli}\ \emph {et~al.}(2019)\citenamefont
  {Mapelli}, \citenamefont {Giacobbo}, \citenamefont {Santoliquido},\ and\
  \citenamefont {Artale}}]{Mapelli:2019bnp}%
  \BibitemOpen
  \bibfield  {author} {\bibinfo {author} {\bibfnamefont {M.}~\bibnamefont
  {Mapelli}}, \bibinfo {author} {\bibfnamefont {N.}~\bibnamefont {Giacobbo}},
  \bibinfo {author} {\bibfnamefont {F.}~\bibnamefont {Santoliquido}},\ and\
  \bibinfo {author} {\bibfnamefont {M.~C.}\ \bibnamefont {Artale}},\ }\bibfield
   {title} {\bibinfo {title} {{The properties of merging black holes and
  neutron stars across cosmic time}},\ }\href
  {https://doi.org/10.1093/mnras/stz1150} {\bibfield  {journal} {\bibinfo
  {journal} {Mon. Not. Roy. Astron. Soc.}\ }\textbf {\bibinfo {volume} {487}},\
  \bibinfo {pages} {2} (\bibinfo {year} {2019})},\ \Eprint
  {https://arxiv.org/abs/1902.01419} {arXiv:1902.01419 [astro-ph.HE]}
  \BibitemShut {NoStop}%
\bibitem [{\citenamefont {Breivik}\ \emph {et~al.}(2020)\citenamefont {Breivik}
  \emph {et~al.}}]{Breivik:2019lmt}%
  \BibitemOpen
  \bibfield  {author} {\bibinfo {author} {\bibfnamefont {K.}~\bibnamefont
  {Breivik}} \emph {et~al.},\ }\bibfield  {title} {\bibinfo {title} {{COSMIC
  Variance in Binary Population Synthesis}},\ }\href
  {https://doi.org/10.3847/1538-4357/ab9d85} {\bibfield  {journal} {\bibinfo
  {journal} {Astrophys. J.}\ }\textbf {\bibinfo {volume} {898}},\ \bibinfo
  {pages} {71} (\bibinfo {year} {2020})},\ \Eprint
  {https://arxiv.org/abs/1911.00903} {arXiv:1911.00903 [astro-ph.HE]}
  \BibitemShut {NoStop}%
\bibitem [{\citenamefont {Bavera}\ \emph {et~al.}(2020)\citenamefont {Bavera},
  \citenamefont {Fragos}, \citenamefont {Qin}, \citenamefont {Zapartas},
  \citenamefont {Neijssel}, \citenamefont {Mandel}, \citenamefont {Batta},
  \citenamefont {Gaebel}, \citenamefont {Kimball},\ and\ \citenamefont
  {Stevenson}}]{Bavera:2019fkg}%
  \BibitemOpen
  \bibfield  {author} {\bibinfo {author} {\bibfnamefont {S.~S.}\ \bibnamefont
  {Bavera}}, \bibinfo {author} {\bibfnamefont {T.}~\bibnamefont {Fragos}},
  \bibinfo {author} {\bibfnamefont {Y.}~\bibnamefont {Qin}}, \bibinfo {author}
  {\bibfnamefont {E.}~\bibnamefont {Zapartas}}, \bibinfo {author}
  {\bibfnamefont {C.~J.}\ \bibnamefont {Neijssel}}, \bibinfo {author}
  {\bibfnamefont {I.}~\bibnamefont {Mandel}}, \bibinfo {author} {\bibfnamefont
  {A.}~\bibnamefont {Batta}}, \bibinfo {author} {\bibfnamefont {S.~M.}\
  \bibnamefont {Gaebel}}, \bibinfo {author} {\bibfnamefont {C.}~\bibnamefont
  {Kimball}},\ and\ \bibinfo {author} {\bibfnamefont {S.}~\bibnamefont
  {Stevenson}},\ }\bibfield  {title} {\bibinfo {title} {{The origin of spin in
  binary black holes: Predicting the distributions of the main observables of
  Advanced LIGO}},\ }\href {https://doi.org/10.1051/0004-6361/201936204}
  {\bibfield  {journal} {\bibinfo  {journal} {Astron. Astrophys.}\ }\textbf
  {\bibinfo {volume} {635}},\ \bibinfo {pages} {A97} (\bibinfo {year}
  {2020})},\ \Eprint {https://arxiv.org/abs/1906.12257} {arXiv:1906.12257
  [astro-ph.HE]} \BibitemShut {NoStop}%
\bibitem [{\citenamefont {Broekgaarden}\ \emph {et~al.}(2019)\citenamefont
  {Broekgaarden}, \citenamefont {Justham}, \citenamefont {de~Mink},
  \citenamefont {Gair}, \citenamefont {Mandel}, \citenamefont {Stevenson},
  \citenamefont {Barrett}, \citenamefont {Vigna-G\'omez},\ and\ \citenamefont
  {Neijssel}}]{Broekgaarden:2019qnw}%
  \BibitemOpen
  \bibfield  {author} {\bibinfo {author} {\bibfnamefont {F.~S.}\ \bibnamefont
  {Broekgaarden}}, \bibinfo {author} {\bibfnamefont {S.}~\bibnamefont
  {Justham}}, \bibinfo {author} {\bibfnamefont {S.~E.}\ \bibnamefont
  {de~Mink}}, \bibinfo {author} {\bibfnamefont {J.}~\bibnamefont {Gair}},
  \bibinfo {author} {\bibfnamefont {I.}~\bibnamefont {Mandel}}, \bibinfo
  {author} {\bibfnamefont {S.}~\bibnamefont {Stevenson}}, \bibinfo {author}
  {\bibfnamefont {J.~W.}\ \bibnamefont {Barrett}}, \bibinfo {author}
  {\bibfnamefont {A.}~\bibnamefont {Vigna-G\'omez}},\ and\ \bibinfo {author}
  {\bibfnamefont {C.~J.}\ \bibnamefont {Neijssel}},\ }\bibfield  {title}
  {\bibinfo {title} {{STROOPWAFEL: Simulating rare outcomes from astrophysical
  populations, with application to gravitational-wave sources}},\ }\href
  {https://doi.org/10.1093/mnras/stz2558} {\bibfield  {journal} {\bibinfo
  {journal} {Mon. Not. Roy. Astron. Soc.}\ }\textbf {\bibinfo {volume} {490}},\
  \bibinfo {pages} {5228} (\bibinfo {year} {2019})},\ \Eprint
  {https://arxiv.org/abs/1905.00910} {arXiv:1905.00910 [astro-ph.HE]}
  \BibitemShut {NoStop}%
\bibitem [{\citenamefont {{Portegies Zwart}}\ and\ \citenamefont
  {{McMillan}}(2000)}]{2000ApJ...528L..17P}%
  \BibitemOpen
  \bibfield  {author} {\bibinfo {author} {\bibfnamefont {S.~F.}\ \bibnamefont
  {{Portegies Zwart}}}\ and\ \bibinfo {author} {\bibfnamefont {S.~L.~W.}\
  \bibnamefont {{McMillan}}},\ }\bibfield  {title} {\bibinfo {title} {{Black
  Hole Mergers in the Universe}},\ }\href {https://doi.org/10.1086/312422}
  {\bibfield  {journal} {\bibinfo  {journal} {\apjl}\ }\textbf {\bibinfo
  {volume} {528}},\ \bibinfo {pages} {L17} (\bibinfo {year} {2000})},\ \Eprint
  {https://arxiv.org/abs/astro-ph/9910061} {arXiv:astro-ph/9910061 [astro-ph]}
  \BibitemShut {NoStop}%
\bibitem [{\citenamefont {Antonini}\ and\ \citenamefont
  {Gieles}(2020)}]{Antonini:2020xnd}%
  \BibitemOpen
  \bibfield  {author} {\bibinfo {author} {\bibfnamefont {F.}~\bibnamefont
  {Antonini}}\ and\ \bibinfo {author} {\bibfnamefont {M.}~\bibnamefont
  {Gieles}},\ }\bibfield  {title} {\bibinfo {title} {{Merger rate of black hole
  binaries from globular clusters: theoretical error bars and comparison to
  gravitational wave data}},\ }\href@noop {} {\bibfield  {journal} {\bibinfo
  {journal} {arXiv e-print}\ } (\bibinfo {year} {2020})},\ \Eprint
  {https://arxiv.org/abs/2009.01861} {arXiv:2009.01861 [astro-ph.HE]}
  \BibitemShut {NoStop}%
\bibitem [{\citenamefont {Santoliquido}\ \emph {et~al.}(2020)\citenamefont
  {Santoliquido}, \citenamefont {Mapelli}, \citenamefont {Giacobbo},
  \citenamefont {Bouffanais},\ and\ \citenamefont
  {Artale}}]{Santoliquido:2020axb}%
  \BibitemOpen
  \bibfield  {author} {\bibinfo {author} {\bibfnamefont {F.}~\bibnamefont
  {Santoliquido}}, \bibinfo {author} {\bibfnamefont {M.}~\bibnamefont
  {Mapelli}}, \bibinfo {author} {\bibfnamefont {N.}~\bibnamefont {Giacobbo}},
  \bibinfo {author} {\bibfnamefont {Y.}~\bibnamefont {Bouffanais}},\ and\
  \bibinfo {author} {\bibfnamefont {M.~C.}\ \bibnamefont {Artale}},\ }\bibfield
   {title} {\bibinfo {title} {{The cosmic merger rate density of compact
  objects: impact of star formation, metallicity, initial mass function and
  binary evolution}},\ }\href@noop {} {\bibfield  {journal} {\bibinfo
  {journal} {arXiv e-print}\ } (\bibinfo {year} {2020})},\ \Eprint
  {https://arxiv.org/abs/2009.03911} {arXiv:2009.03911 [astro-ph.HE]}
  \BibitemShut {NoStop}%
\bibitem [{\citenamefont {Rodriguez}\ \emph {et~al.}(2015)\citenamefont
  {Rodriguez}, \citenamefont {Morscher}, \citenamefont {Pattabiraman},
  \citenamefont {Chatterjee}, \citenamefont {Haster},\ and\ \citenamefont
  {Rasio}}]{Rodriguez:2015oxa}%
  \BibitemOpen
  \bibfield  {author} {\bibinfo {author} {\bibfnamefont {C.~L.}\ \bibnamefont
  {Rodriguez}}, \bibinfo {author} {\bibfnamefont {M.}~\bibnamefont {Morscher}},
  \bibinfo {author} {\bibfnamefont {B.}~\bibnamefont {Pattabiraman}}, \bibinfo
  {author} {\bibfnamefont {S.}~\bibnamefont {Chatterjee}}, \bibinfo {author}
  {\bibfnamefont {C.-J.}\ \bibnamefont {Haster}},\ and\ \bibinfo {author}
  {\bibfnamefont {F.~A.}\ \bibnamefont {Rasio}},\ }\bibfield  {title} {\bibinfo
  {title} {{Binary Black Hole Mergers from Globular Clusters: Implications for
  Advanced LIGO}},\ }\href {https://doi.org/10.1103/PhysRevLett.115.051101}
  {\bibfield  {journal} {\bibinfo  {journal} {Phys. Rev. Lett.}\ }\textbf
  {\bibinfo {volume} {115}},\ \bibinfo {pages} {051101} (\bibinfo {year}
  {2015})},\ \bibinfo {note} {[Erratum: Phys.Rev.Lett. 116, 029901 (2016)]},\
  \Eprint {https://arxiv.org/abs/1505.00792} {arXiv:1505.00792 [astro-ph.HE]}
  \BibitemShut {NoStop}%
\bibitem [{\citenamefont {Rodriguez}\ \emph {et~al.}(2016)\citenamefont
  {Rodriguez}, \citenamefont {Chatterjee},\ and\ \citenamefont
  {Rasio}}]{Rodriguez:2016kxx}%
  \BibitemOpen
  \bibfield  {author} {\bibinfo {author} {\bibfnamefont {C.~L.}\ \bibnamefont
  {Rodriguez}}, \bibinfo {author} {\bibfnamefont {S.}~\bibnamefont
  {Chatterjee}},\ and\ \bibinfo {author} {\bibfnamefont {F.~A.}\ \bibnamefont
  {Rasio}},\ }\bibfield  {title} {\bibinfo {title} {{Binary Black Hole Mergers
  from Globular Clusters: Masses, Merger Rates, and the Impact of Stellar
  Evolution}},\ }\href {https://doi.org/10.1103/PhysRevD.93.084029} {\bibfield
  {journal} {\bibinfo  {journal} {Phys. Rev. D}\ }\textbf {\bibinfo {volume}
  {93}},\ \bibinfo {pages} {084029} (\bibinfo {year} {2016})},\ \Eprint
  {https://arxiv.org/abs/1602.02444} {arXiv:1602.02444 [astro-ph.HE]}
  \BibitemShut {NoStop}%
\bibitem [{\citenamefont {Rodriguez}\ and\ \citenamefont
  {Loeb}(2018)}]{Rodriguez:2018rmd}%
  \BibitemOpen
  \bibfield  {author} {\bibinfo {author} {\bibfnamefont {C.~L.}\ \bibnamefont
  {Rodriguez}}\ and\ \bibinfo {author} {\bibfnamefont {A.}~\bibnamefont
  {Loeb}},\ }\bibfield  {title} {\bibinfo {title} {{Redshift Evolution of the
  Black Hole Merger Rate from Globular Clusters}},\ }\href
  {https://doi.org/10.3847/2041-8213/aae377} {\bibfield  {journal} {\bibinfo
  {journal} {Astrophys. J. Lett.}\ }\textbf {\bibinfo {volume} {866}},\
  \bibinfo {pages} {L5} (\bibinfo {year} {2018})},\ \Eprint
  {https://arxiv.org/abs/1809.01152} {arXiv:1809.01152 [astro-ph.HE]}
  \BibitemShut {NoStop}%
\bibitem [{\citenamefont {Di~Carlo}\ \emph {et~al.}(2019)\citenamefont
  {Di~Carlo}, \citenamefont {Giacobbo}, \citenamefont {Mapelli}, \citenamefont
  {Pasquato}, \citenamefont {Spera}, \citenamefont {Wang},\ and\ \citenamefont
  {Haardt}}]{DiCarlo:2019pmf}%
  \BibitemOpen
  \bibfield  {author} {\bibinfo {author} {\bibfnamefont {U.~N.}\ \bibnamefont
  {Di~Carlo}}, \bibinfo {author} {\bibfnamefont {N.}~\bibnamefont {Giacobbo}},
  \bibinfo {author} {\bibfnamefont {M.}~\bibnamefont {Mapelli}}, \bibinfo
  {author} {\bibfnamefont {M.}~\bibnamefont {Pasquato}}, \bibinfo {author}
  {\bibfnamefont {M.}~\bibnamefont {Spera}}, \bibinfo {author} {\bibfnamefont
  {L.}~\bibnamefont {Wang}},\ and\ \bibinfo {author} {\bibfnamefont
  {F.}~\bibnamefont {Haardt}},\ }\bibfield  {title} {\bibinfo {title} {{Merging
  black holes in young star clusters}},\ }\href
  {https://doi.org/10.1093/mnras/stz1453} {\bibfield  {journal} {\bibinfo
  {journal} {Mon. Not. Roy. Astron. Soc.}\ }\textbf {\bibinfo {volume} {487}},\
  \bibinfo {pages} {2947} (\bibinfo {year} {2019})},\ \Eprint
  {https://arxiv.org/abs/1901.00863} {arXiv:1901.00863 [astro-ph.HE]}
  \BibitemShut {NoStop}%
\bibitem [{\citenamefont {Kremer}\ \emph {et~al.}(2020)\citenamefont {Kremer},
  \citenamefont {Spera}, \citenamefont {Becker}, \citenamefont {Chatterjee},
  \citenamefont {Di~Carlo}, \citenamefont {Fragione}, \citenamefont
  {Rodriguez}, \citenamefont {Ye},\ and\ \citenamefont
  {Rasio}}]{Kremer:2020wtp}%
  \BibitemOpen
  \bibfield  {author} {\bibinfo {author} {\bibfnamefont {K.}~\bibnamefont
  {Kremer}}, \bibinfo {author} {\bibfnamefont {M.}~\bibnamefont {Spera}},
  \bibinfo {author} {\bibfnamefont {D.}~\bibnamefont {Becker}}, \bibinfo
  {author} {\bibfnamefont {S.}~\bibnamefont {Chatterjee}}, \bibinfo {author}
  {\bibfnamefont {U.~N.}\ \bibnamefont {Di~Carlo}}, \bibinfo {author}
  {\bibfnamefont {G.}~\bibnamefont {Fragione}}, \bibinfo {author}
  {\bibfnamefont {C.~L.}\ \bibnamefont {Rodriguez}}, \bibinfo {author}
  {\bibfnamefont {C.~S.}\ \bibnamefont {Ye}},\ and\ \bibinfo {author}
  {\bibfnamefont {F.~A.}\ \bibnamefont {Rasio}},\ }\bibfield  {title} {\bibinfo
  {title} {{Populating the upper black hole mass gap through stellar collisions
  in young star clusters}},\ }\href {https://doi.org/10.3847/1538-4357/abb945}
  {\bibfield  {journal} {\bibinfo  {journal} {Astrophys. J.}\ }\textbf
  {\bibinfo {volume} {903}},\ \bibinfo {pages} {45} (\bibinfo {year} {2020})},\
  \Eprint {https://arxiv.org/abs/2006.10771} {arXiv:2006.10771 [astro-ph.HE]}
  \BibitemShut {NoStop}%
\bibitem [{\citenamefont {Bartos}\ \emph {et~al.}(2017)\citenamefont {Bartos},
  \citenamefont {Kocsis}, \citenamefont {Haiman},\ and\ \citenamefont
  {M\'arka}}]{Bartos:2016dgn}%
  \BibitemOpen
  \bibfield  {author} {\bibinfo {author} {\bibfnamefont {I.}~\bibnamefont
  {Bartos}}, \bibinfo {author} {\bibfnamefont {B.}~\bibnamefont {Kocsis}},
  \bibinfo {author} {\bibfnamefont {Z.}~\bibnamefont {Haiman}},\ and\ \bibinfo
  {author} {\bibfnamefont {S.}~\bibnamefont {M\'arka}},\ }\bibfield  {title}
  {\bibinfo {title} {{Rapid and Bright Stellar-mass Binary Black Hole Mergers
  in Active Galactic Nuclei}},\ }\href
  {https://doi.org/10.3847/1538-4357/835/2/165} {\bibfield  {journal} {\bibinfo
   {journal} {Astrophys. J.}\ }\textbf {\bibinfo {volume} {835}},\ \bibinfo
  {pages} {165} (\bibinfo {year} {2017})},\ \Eprint
  {https://arxiv.org/abs/1602.03831} {arXiv:1602.03831 [astro-ph.HE]}
  \BibitemShut {NoStop}%
\bibitem [{\citenamefont {Yi}\ and\ \citenamefont {Cheng}(2019)}]{Yi:2019rwo}%
  \BibitemOpen
  \bibfield  {author} {\bibinfo {author} {\bibfnamefont {S.-X.}\ \bibnamefont
  {Yi}}\ and\ \bibinfo {author} {\bibfnamefont {K.}~\bibnamefont {Cheng}},\
  }\bibfield  {title} {\bibinfo {title} {{Where Are the Electromagnetic-wave
  Counterparts of Stellar-mass Binary Black Hole Mergers?}},\ }\href
  {https://doi.org/10.3847/2041-8213/ab459a} {\bibfield  {journal} {\bibinfo
  {journal} {Astrophys. J. Lett.}\ }\textbf {\bibinfo {volume} {884}},\
  \bibinfo {pages} {L12} (\bibinfo {year} {2019})},\ \Eprint
  {https://arxiv.org/abs/1909.08384} {arXiv:1909.08384 [astro-ph.HE]}
  \BibitemShut {NoStop}%
\bibitem [{\citenamefont {Yang}\ \emph {et~al.}(2019)\citenamefont {Yang} \emph
  {et~al.}}]{Yang:2019cbr}%
  \BibitemOpen
  \bibfield  {author} {\bibinfo {author} {\bibfnamefont {Y.}~\bibnamefont
  {Yang}} \emph {et~al.},\ }\bibfield  {title} {\bibinfo {title} {{Hierarchical
  Black Hole Mergers in Active Galactic Nuclei}},\ }\href
  {https://doi.org/10.1103/PhysRevLett.123.181101} {\bibfield  {journal}
  {\bibinfo  {journal} {Phys. Rev. Lett.}\ }\textbf {\bibinfo {volume} {123}},\
  \bibinfo {pages} {181101} (\bibinfo {year} {2019})},\ \Eprint
  {https://arxiv.org/abs/1906.09281} {arXiv:1906.09281 [astro-ph.HE]}
  \BibitemShut {NoStop}%
\bibitem [{\citenamefont {Yang}\ \emph {et~al.}(2020)\citenamefont {Yang},
  \citenamefont {Bartos}, \citenamefont {Haiman}, \citenamefont {Kocsis},
  \citenamefont {M\'arka},\ and\ \citenamefont {Tagawa}}]{Yang:2020lhq}%
  \BibitemOpen
  \bibfield  {author} {\bibinfo {author} {\bibfnamefont {Y.}~\bibnamefont
  {Yang}}, \bibinfo {author} {\bibfnamefont {I.}~\bibnamefont {Bartos}},
  \bibinfo {author} {\bibfnamefont {Z.}~\bibnamefont {Haiman}}, \bibinfo
  {author} {\bibfnamefont {B.}~\bibnamefont {Kocsis}}, \bibinfo {author}
  {\bibfnamefont {S.}~\bibnamefont {M\'arka}},\ and\ \bibinfo {author}
  {\bibfnamefont {H.}~\bibnamefont {Tagawa}},\ }\bibfield  {title} {\bibinfo
  {title} {{Cosmic Evolution of Stellar-mass Black Hole Merger Rate in Active
  Galactic Nuclei}},\ }\href {https://doi.org/10.3847/1538-4357/ab91b4}
  {\bibfield  {journal} {\bibinfo  {journal} {Astrophys. J.}\ }\textbf
  {\bibinfo {volume} {896}},\ \bibinfo {pages} {138} (\bibinfo {year}
  {2020})},\ \Eprint {https://arxiv.org/abs/2003.08564} {arXiv:2003.08564
  [astro-ph.HE]} \BibitemShut {NoStop}%
\bibitem [{\citenamefont {Gr\"obner}\ \emph {et~al.}(2020)\citenamefont
  {Gr\"obner}, \citenamefont {Ishibashi}, \citenamefont {Tiwari}, \citenamefont
  {Haney},\ and\ \citenamefont {Jetzer}}]{Grobner:2020drr}%
  \BibitemOpen
  \bibfield  {author} {\bibinfo {author} {\bibfnamefont {M.}~\bibnamefont
  {Gr\"obner}}, \bibinfo {author} {\bibfnamefont {W.}~\bibnamefont
  {Ishibashi}}, \bibinfo {author} {\bibfnamefont {S.}~\bibnamefont {Tiwari}},
  \bibinfo {author} {\bibfnamefont {M.}~\bibnamefont {Haney}},\ and\ \bibinfo
  {author} {\bibfnamefont {P.}~\bibnamefont {Jetzer}},\ }\bibfield  {title}
  {\bibinfo {title} {{Binary black hole mergers in AGN accretion discs:
  gravitational wave rate density estimates}},\ }\href
  {https://doi.org/10.1051/0004-6361/202037681} {\bibfield  {journal} {\bibinfo
   {journal} {Astron. Astrophys.}\ }\textbf {\bibinfo {volume} {638}},\
  \bibinfo {pages} {A119} (\bibinfo {year} {2020})},\ \Eprint
  {https://arxiv.org/abs/2005.03571} {arXiv:2005.03571 [astro-ph.GA]}
  \BibitemShut {NoStop}%
\bibitem [{\citenamefont {Tagawa}\ \emph
  {et~al.}(2020{\natexlab{a}})\citenamefont {Tagawa}, \citenamefont {Haiman},\
  and\ \citenamefont {Kocsis}}]{Tagawa:2019osr}%
  \BibitemOpen
  \bibfield  {author} {\bibinfo {author} {\bibfnamefont {H.}~\bibnamefont
  {Tagawa}}, \bibinfo {author} {\bibfnamefont {Z.}~\bibnamefont {Haiman}},\
  and\ \bibinfo {author} {\bibfnamefont {B.}~\bibnamefont {Kocsis}},\
  }\bibfield  {title} {\bibinfo {title} {{Formation and Evolution of Compact
  Object Binaries in AGN Disks}},\ }\href
  {https://doi.org/10.3847/1538-4357/ab9b8c} {\bibfield  {journal} {\bibinfo
  {journal} {Astrophys. J.}\ }\textbf {\bibinfo {volume} {898}},\ \bibinfo
  {pages} {25} (\bibinfo {year} {2020}{\natexlab{a}})},\ \Eprint
  {https://arxiv.org/abs/1912.08218} {arXiv:1912.08218 [astro-ph.GA]}
  \BibitemShut {NoStop}%
\bibitem [{\citenamefont {Tagawa}\ \emph
  {et~al.}(2020{\natexlab{b}})\citenamefont {Tagawa}, \citenamefont {Kocsis},
  \citenamefont {Haiman}, \citenamefont {Bartos}, \citenamefont {Omukai},\ and\
  \citenamefont {Samsing}}]{Tagawa:2020qll}%
  \BibitemOpen
  \bibfield  {author} {\bibinfo {author} {\bibfnamefont {H.}~\bibnamefont
  {Tagawa}}, \bibinfo {author} {\bibfnamefont {B.}~\bibnamefont {Kocsis}},
  \bibinfo {author} {\bibfnamefont {Z.}~\bibnamefont {Haiman}}, \bibinfo
  {author} {\bibfnamefont {I.}~\bibnamefont {Bartos}}, \bibinfo {author}
  {\bibfnamefont {K.}~\bibnamefont {Omukai}},\ and\ \bibinfo {author}
  {\bibfnamefont {J.}~\bibnamefont {Samsing}},\ }\bibfield  {title} {\bibinfo
  {title} {{Mass-gap Mergers in Active Galactic Nuclei}},\ }\href@noop {}
  {\bibfield  {journal} {\bibinfo  {journal} {arXiv e-print}\ } (\bibinfo
  {year} {2020}{\natexlab{b}})},\ \Eprint {https://arxiv.org/abs/2012.00011}
  {arXiv:2012.00011 [astro-ph.HE]} \BibitemShut {NoStop}%
\bibitem [{\citenamefont {Tagawa}\ \emph
  {et~al.}(2020{\natexlab{c}})\citenamefont {Tagawa}, \citenamefont {Haiman},
  \citenamefont {Bartos},\ and\ \citenamefont {Kocsis}}]{Tagawa:2020dxe}%
  \BibitemOpen
  \bibfield  {author} {\bibinfo {author} {\bibfnamefont {H.}~\bibnamefont
  {Tagawa}}, \bibinfo {author} {\bibfnamefont {Z.}~\bibnamefont {Haiman}},
  \bibinfo {author} {\bibfnamefont {I.}~\bibnamefont {Bartos}},\ and\ \bibinfo
  {author} {\bibfnamefont {B.}~\bibnamefont {Kocsis}},\ }\bibfield  {title}
  {\bibinfo {title} {{Spin Evolution of Stellar-mass Black Hole Binaries in
  Active Galactic Nuclei}},\ }\href {https://doi.org/10.3847/1538-4357/aba2cc}
  {\bibfield  {journal} {\bibinfo  {journal} {Astrophys. J.}\ }\textbf
  {\bibinfo {volume} {899}},\ \bibinfo {pages} {26} (\bibinfo {year}
  {2020}{\natexlab{c}})},\ \Eprint {https://arxiv.org/abs/2004.11914}
  {arXiv:2004.11914 [astro-ph.HE]} \BibitemShut {NoStop}%
\bibitem [{\citenamefont {Samsing}\ \emph {et~al.}(2020)\citenamefont
  {Samsing}, \citenamefont {Bartos}, \citenamefont {D'Orazio}, \citenamefont
  {Haiman}, \citenamefont {Kocsis}, \citenamefont {Leigh}, \citenamefont {Liu},
  \citenamefont {Pessah},\ and\ \citenamefont {Tagawa}}]{Samsing:2020tda}%
  \BibitemOpen
  \bibfield  {author} {\bibinfo {author} {\bibfnamefont {J.}~\bibnamefont
  {Samsing}}, \bibinfo {author} {\bibfnamefont {I.}~\bibnamefont {Bartos}},
  \bibinfo {author} {\bibfnamefont {D.}~\bibnamefont {D'Orazio}}, \bibinfo
  {author} {\bibfnamefont {Z.}~\bibnamefont {Haiman}}, \bibinfo {author}
  {\bibfnamefont {B.}~\bibnamefont {Kocsis}}, \bibinfo {author} {\bibfnamefont
  {N.}~\bibnamefont {Leigh}}, \bibinfo {author} {\bibfnamefont
  {B.}~\bibnamefont {Liu}}, \bibinfo {author} {\bibfnamefont {M.}~\bibnamefont
  {Pessah}},\ and\ \bibinfo {author} {\bibfnamefont {H.}~\bibnamefont
  {Tagawa}},\ }\bibfield  {title} {\bibinfo {title} {{Active Galactic Nuclei as
  Factories for Eccentric Black Hole Mergers}},\ }\href@noop {} {\bibfield
  {journal} {\bibinfo  {journal} {Arxiv e-print}\ } (\bibinfo {year} {2020})},\
  \Eprint {https://arxiv.org/abs/2010.09765} {arXiv:2010.09765 [astro-ph.HE]}
  \BibitemShut {NoStop}%
\bibitem [{\citenamefont {Kinugawa}\ \emph {et~al.}(2014)\citenamefont
  {Kinugawa}, \citenamefont {Inayoshi}, \citenamefont {Hotokezaka},
  \citenamefont {Nakauchi},\ and\ \citenamefont {Nakamura}}]{Kinugawa:2014zha}%
  \BibitemOpen
  \bibfield  {author} {\bibinfo {author} {\bibfnamefont {T.}~\bibnamefont
  {Kinugawa}}, \bibinfo {author} {\bibfnamefont {K.}~\bibnamefont {Inayoshi}},
  \bibinfo {author} {\bibfnamefont {K.}~\bibnamefont {Hotokezaka}}, \bibinfo
  {author} {\bibfnamefont {D.}~\bibnamefont {Nakauchi}},\ and\ \bibinfo
  {author} {\bibfnamefont {T.}~\bibnamefont {Nakamura}},\ }\bibfield  {title}
  {\bibinfo {title} {{Possible Indirect Confirmation of the Existence of Pop
  III Massive Stars by Gravitational Wave}},\ }\href
  {https://doi.org/10.1093/mnras/stu1022} {\bibfield  {journal} {\bibinfo
  {journal} {Mon. Not. Roy. Astron. Soc.}\ }\textbf {\bibinfo {volume} {442}},\
  \bibinfo {pages} {2963} (\bibinfo {year} {2014})},\ \Eprint
  {https://arxiv.org/abs/1402.6672} {arXiv:1402.6672 [astro-ph.HE]}
  \BibitemShut {NoStop}%
\bibitem [{\citenamefont {Kinugawa}\ \emph {et~al.}(2016)\citenamefont
  {Kinugawa}, \citenamefont {Miyamoto}, \citenamefont {Kanda},\ and\
  \citenamefont {Nakamura}}]{Kinugawa:2015nla}%
  \BibitemOpen
  \bibfield  {author} {\bibinfo {author} {\bibfnamefont {T.}~\bibnamefont
  {Kinugawa}}, \bibinfo {author} {\bibfnamefont {A.}~\bibnamefont {Miyamoto}},
  \bibinfo {author} {\bibfnamefont {N.}~\bibnamefont {Kanda}},\ and\ \bibinfo
  {author} {\bibfnamefont {T.}~\bibnamefont {Nakamura}},\ }\bibfield  {title}
  {\bibinfo {title} {{The detection rate of inspiral and quasi-normal modes of
  Population III binary black holes which can confirm or refute the general
  relativity in the strong gravity region}},\ }\href
  {https://doi.org/10.1093/mnras/stv2624} {\bibfield  {journal} {\bibinfo
  {journal} {Mon. Not. Roy. Astron. Soc.}\ }\textbf {\bibinfo {volume} {456}},\
  \bibinfo {pages} {1093} (\bibinfo {year} {2016})},\ \Eprint
  {https://arxiv.org/abs/1505.06962} {arXiv:1505.06962 [astro-ph.SR]}
  \BibitemShut {NoStop}%
\bibitem [{\citenamefont {Hartwig}\ \emph {et~al.}(2016)\citenamefont
  {Hartwig}, \citenamefont {Volonteri}, \citenamefont {Bromm}, \citenamefont
  {Klessen}, \citenamefont {Barausse}, \citenamefont {Magg},\ and\
  \citenamefont {Stacy}}]{Hartwig:2016nde}%
  \BibitemOpen
  \bibfield  {author} {\bibinfo {author} {\bibfnamefont {T.}~\bibnamefont
  {Hartwig}}, \bibinfo {author} {\bibfnamefont {M.}~\bibnamefont {Volonteri}},
  \bibinfo {author} {\bibfnamefont {V.}~\bibnamefont {Bromm}}, \bibinfo
  {author} {\bibfnamefont {R.~S.}\ \bibnamefont {Klessen}}, \bibinfo {author}
  {\bibfnamefont {E.}~\bibnamefont {Barausse}}, \bibinfo {author}
  {\bibfnamefont {M.}~\bibnamefont {Magg}},\ and\ \bibinfo {author}
  {\bibfnamefont {A.}~\bibnamefont {Stacy}},\ }\bibfield  {title} {\bibinfo
  {title} {{Gravitational Waves from the Remnants of the First Stars}},\ }\href
  {https://doi.org/10.1093/mnrasl/slw074} {\bibfield  {journal} {\bibinfo
  {journal} {Mon. Not. Roy. Astron. Soc.}\ }\textbf {\bibinfo {volume} {460}},\
  \bibinfo {pages} {L74} (\bibinfo {year} {2016})},\ \Eprint
  {https://arxiv.org/abs/1603.05655} {arXiv:1603.05655 [astro-ph.GA]}
  \BibitemShut {NoStop}%
\bibitem [{\citenamefont {Belczynski}\ \emph {et~al.}(2017)\citenamefont
  {Belczynski}, \citenamefont {Ryu}, \citenamefont {Perna}, \citenamefont
  {Berti}, \citenamefont {Tanaka},\ and\ \citenamefont
  {Bulik}}]{Belczynski:2016ieo}%
  \BibitemOpen
  \bibfield  {author} {\bibinfo {author} {\bibfnamefont {K.}~\bibnamefont
  {Belczynski}}, \bibinfo {author} {\bibfnamefont {T.}~\bibnamefont {Ryu}},
  \bibinfo {author} {\bibfnamefont {R.}~\bibnamefont {Perna}}, \bibinfo
  {author} {\bibfnamefont {E.}~\bibnamefont {Berti}}, \bibinfo {author}
  {\bibfnamefont {T.}~\bibnamefont {Tanaka}},\ and\ \bibinfo {author}
  {\bibfnamefont {T.}~\bibnamefont {Bulik}},\ }\bibfield  {title} {\bibinfo
  {title} {{On the likelihood of detecting gravitational waves from Population
  III compact object binaries}},\ }\href
  {https://doi.org/10.1093/mnras/stx1759} {\bibfield  {journal} {\bibinfo
  {journal} {Mon. Not. Roy. Astron. Soc.}\ }\textbf {\bibinfo {volume} {471}},\
  \bibinfo {pages} {4702} (\bibinfo {year} {2017})},\ \Eprint
  {https://arxiv.org/abs/1612.01524} {arXiv:1612.01524 [astro-ph.HE]}
  \BibitemShut {NoStop}%
\bibitem [{\citenamefont {De~Luca}\ \emph
  {et~al.}(2021{\natexlab{a}})\citenamefont {De~Luca}, \citenamefont
  {Franciolini}, \citenamefont {Pani},\ and\ \citenamefont
  {Riotto}}]{DeLuca:2021hde}%
  \BibitemOpen
  \bibfield  {author} {\bibinfo {author} {\bibfnamefont {V.}~\bibnamefont
  {De~Luca}}, \bibinfo {author} {\bibfnamefont {G.}~\bibnamefont
  {Franciolini}}, \bibinfo {author} {\bibfnamefont {P.}~\bibnamefont {Pani}},\
  and\ \bibinfo {author} {\bibfnamefont {A.}~\bibnamefont {Riotto}},\
  }\bibfield  {title} {\bibinfo {title} {{The Minimum Testable Abundance of
  Primordial Black Holes at Future Gravitational-Wave Detectors}},\ }\href@noop
  {} {\bibfield  {journal} {\bibinfo  {journal} {ArXiv}\ } (\bibinfo {year}
  {2021}{\natexlab{a}})},\ \Eprint {https://arxiv.org/abs/2106.13769}
  {arXiv:2106.13769 [astro-ph.CO]} \BibitemShut {NoStop}%
\bibitem [{\citenamefont {De~Luca}\ \emph
  {et~al.}(2021{\natexlab{b}})\citenamefont {De~Luca}, \citenamefont
  {Franciolini}, \citenamefont {Pani},\ and\ \citenamefont
  {Riotto}}]{DeLuca:2021wjr}%
  \BibitemOpen
  \bibfield  {author} {\bibinfo {author} {\bibfnamefont {V.}~\bibnamefont
  {De~Luca}}, \bibinfo {author} {\bibfnamefont {G.}~\bibnamefont
  {Franciolini}}, \bibinfo {author} {\bibfnamefont {P.}~\bibnamefont {Pani}},\
  and\ \bibinfo {author} {\bibfnamefont {A.}~\bibnamefont {Riotto}},\
  }\bibfield  {title} {\bibinfo {title} {{Bayesian Evidence for Both
  Astrophysical and Primordial Black Holes: Mapping the GWTC-2 Catalog to
  Third-Generation Detectors}},\ }\href
  {https://doi.org/10.1088/1475-7516/2021/05/003} {\bibfield  {journal}
  {\bibinfo  {journal} {JCAP}\ }\textbf {\bibinfo {volume} {05}},\ \bibinfo
  {pages} {003}},\ \Eprint {https://arxiv.org/abs/2102.03809} {arXiv:2102.03809
  [astro-ph.CO]} \BibitemShut {NoStop}%
\bibitem [{\citenamefont {Abbott}\ \emph {et~al.}(2017)\citenamefont {Abbott}
  \emph {et~al.}}]{Evans:2016mbw}%
  \BibitemOpen
  \bibfield  {author} {\bibinfo {author} {\bibfnamefont {B.~P.}\ \bibnamefont
  {Abbott}} \emph {et~al.} (\bibinfo {collaboration} {LIGO Scientific}),\
  }\bibfield  {title} {\bibinfo {title} {{Exploring the Sensitivity of Next
  Generation Gravitational Wave Detectors}},\ }\href
  {https://doi.org/10.1088/1361-6382/aa51f4} {\bibfield  {journal} {\bibinfo
  {journal} {Class. Quant. Grav.}\ }\textbf {\bibinfo {volume} {34}},\ \bibinfo
  {pages} {044001} (\bibinfo {year} {2017})},\ \Eprint
  {https://arxiv.org/abs/1607.08697} {arXiv:1607.08697 [astro-ph.IM]}
  \BibitemShut {NoStop}%
\bibitem [{\citenamefont {Reitze}\ \emph {et~al.}(2019)\citenamefont {Reitze}
  \emph {et~al.}}]{Reitze:2019iox}%
  \BibitemOpen
  \bibfield  {author} {\bibinfo {author} {\bibfnamefont {D.}~\bibnamefont
  {Reitze}} \emph {et~al.},\ }\bibfield  {title} {\bibinfo {title} {{Cosmic
  Explorer: The U.S. Contribution to Gravitational-Wave Astronomy beyond
  LIGO}},\ }\href@noop {} {\bibfield  {journal} {\bibinfo  {journal} {Bull. Am.
  Astron. Soc.}\ }\textbf {\bibinfo {volume} {51}},\ \bibinfo {pages} {035}
  (\bibinfo {year} {2019})},\ \Eprint {https://arxiv.org/abs/1907.04833}
  {arXiv:1907.04833 [astro-ph.IM]} \BibitemShut {NoStop}%
\bibitem [{\citenamefont {Evans}\ \emph {et~al.}(2021)\citenamefont {Evans}
  \emph {et~al.}}]{CEHS}%
  \BibitemOpen
  \bibfield  {author} {\bibinfo {author} {\bibfnamefont {M.}~\bibnamefont
  {Evans}} \emph {et~al.},\ }\href {dcc.cosmicexplorer.org/CE-P2100003/public}
  {\bibinfo {title} {A horizon study for cosmic explorer science,
  observatories, and community}},\ \bibinfo {howpublished}
  {\url{dcc.cosmicexplorer.org/CE-P2100003/public}} (\bibinfo {year}
  {2021})\BibitemShut {NoStop}%
\bibitem [{\citenamefont {Punturo}\ \emph {et~al.}(2010)\citenamefont {Punturo}
  \emph {et~al.}}]{Punturo:2010zz}%
  \BibitemOpen
  \bibfield  {author} {\bibinfo {author} {\bibfnamefont {M.}~\bibnamefont
  {Punturo}} \emph {et~al.},\ }\bibfield  {title} {\bibinfo {title} {{The
  Einstein Telescope: A third-generation gravitational wave observatory}},\
  }\href {https://doi.org/10.1088/0264-9381/27/19/194002} {\bibfield  {journal}
  {\bibinfo  {journal} {Class. Quant. Grav.}\ }\textbf {\bibinfo {volume}
  {27}},\ \bibinfo {pages} {194002} (\bibinfo {year} {2010})}\BibitemShut
  {NoStop}%
\bibitem [{\citenamefont {Maggiore}\ \emph {et~al.}(2020)\citenamefont
  {Maggiore} \emph {et~al.}}]{Maggiore:2019uih}%
  \BibitemOpen
  \bibfield  {author} {\bibinfo {author} {\bibfnamefont {M.}~\bibnamefont
  {Maggiore}} \emph {et~al.},\ }\bibfield  {title} {\bibinfo {title} {{Science
  Case for the Einstein Telescope}},\ }\href
  {https://doi.org/10.1088/1475-7516/2020/03/050} {\bibfield  {journal}
  {\bibinfo  {journal} {JCAP}\ }\textbf {\bibinfo {volume} {03}},\ \bibinfo
  {pages} {050}},\ \Eprint {https://arxiv.org/abs/1912.02622} {arXiv:1912.02622
  [astro-ph.CO]} \BibitemShut {NoStop}%
\bibitem [{\citenamefont {Mirbabayi}\ \emph {et~al.}(2020)\citenamefont
  {Mirbabayi}, \citenamefont {Gruzinov},\ and\ \citenamefont
  {Nore\~na}}]{Mirbabayi:2019uph}%
  \BibitemOpen
  \bibfield  {author} {\bibinfo {author} {\bibfnamefont {M.}~\bibnamefont
  {Mirbabayi}}, \bibinfo {author} {\bibfnamefont {A.}~\bibnamefont
  {Gruzinov}},\ and\ \bibinfo {author} {\bibfnamefont {J.}~\bibnamefont
  {Nore\~na}},\ }\bibfield  {title} {\bibinfo {title} {{Spin of Primordial
  Black Holes}},\ }\href {https://doi.org/10.1088/1475-7516/2020/03/017}
  {\bibfield  {journal} {\bibinfo  {journal} {JCAP}\ }\textbf {\bibinfo
  {volume} {03}},\ \bibinfo {pages} {017}},\ \Eprint
  {https://arxiv.org/abs/1901.05963} {arXiv:1901.05963 [astro-ph.CO]}
  \BibitemShut {NoStop}%
\bibitem [{\citenamefont {De~Luca}\ \emph {et~al.}(2019)\citenamefont
  {De~Luca}, \citenamefont {Desjacques}, \citenamefont {Franciolini},
  \citenamefont {Malhotra},\ and\ \citenamefont {Riotto}}]{DeLuca:2019buf}%
  \BibitemOpen
  \bibfield  {author} {\bibinfo {author} {\bibfnamefont {V.}~\bibnamefont
  {De~Luca}}, \bibinfo {author} {\bibfnamefont {V.}~\bibnamefont {Desjacques}},
  \bibinfo {author} {\bibfnamefont {G.}~\bibnamefont {Franciolini}}, \bibinfo
  {author} {\bibfnamefont {A.}~\bibnamefont {Malhotra}},\ and\ \bibinfo
  {author} {\bibfnamefont {A.}~\bibnamefont {Riotto}},\ }\bibfield  {title}
  {\bibinfo {title} {{The initial spin probability distribution of primordial
  black holes}},\ }\href {https://doi.org/10.1088/1475-7516/2019/05/018}
  {\bibfield  {journal} {\bibinfo  {journal} {JCAP}\ }\textbf {\bibinfo
  {volume} {05}},\ \bibinfo {pages} {018}},\ \Eprint
  {https://arxiv.org/abs/1903.01179} {arXiv:1903.01179 [astro-ph.CO]}
  \BibitemShut {NoStop}%
\bibitem [{\citenamefont {Bianchi}\ \emph {et~al.}(2018)\citenamefont
  {Bianchi}, \citenamefont {Gupta}, \citenamefont {Haggard},\ and\
  \citenamefont {Sathyaprakash}}]{Bianchi:2018ula}%
  \BibitemOpen
  \bibfield  {author} {\bibinfo {author} {\bibfnamefont {E.}~\bibnamefont
  {Bianchi}}, \bibinfo {author} {\bibfnamefont {A.}~\bibnamefont {Gupta}},
  \bibinfo {author} {\bibfnamefont {H.~M.}\ \bibnamefont {Haggard}},\ and\
  \bibinfo {author} {\bibfnamefont {B.~S.}\ \bibnamefont {Sathyaprakash}},\
  }\bibfield  {title} {\bibinfo {title} {{Small Spins of Primordial Black Holes
  from Random Geometries: Bekenstein-Hawking Entropy and Gravitational Wave
  Observations}},\ }\href@noop {} {\bibfield  {journal} {\bibinfo  {journal}
  {arXiv}\ } (\bibinfo {year} {2018})},\ \Eprint
  {https://arxiv.org/abs/1812.05127} {arXiv:1812.05127 [gr-qc]} \BibitemShut
  {NoStop}%
\bibitem [{Note1()}]{Note1}%
  \BibitemOpen
  \bibinfo {note} {Throughout this study, we use a $\Lambda $CDM cosmology
  based on the Planck 2018 results~\cite {Planck:2018vyg} to convert luminosity
  distance into redshift.}\BibitemShut {Stop}%
\bibitem [{\citenamefont {Skilling}(2006)}]{Skilling:2006gxv}%
  \BibitemOpen
  \bibfield  {author} {\bibinfo {author} {\bibfnamefont {J.}~\bibnamefont
  {Skilling}},\ }\bibfield  {title} {\bibinfo {title} {{Nested sampling for
  general Bayesian computation}},\ }\href {https://doi.org/10.1214/06-BA127}
  {\bibfield  {journal} {\bibinfo  {journal} {Bayesian Analysis}\ }\textbf
  {\bibinfo {volume} {1}},\ \bibinfo {pages} {833} (\bibinfo {year}
  {2006})}\BibitemShut {NoStop}%
\bibitem [{\citenamefont {{Speagle}}(2020)}]{2020MNRAS.493.3132S}%
  \BibitemOpen
  \bibfield  {author} {\bibinfo {author} {\bibfnamefont {J.~S.}\ \bibnamefont
  {{Speagle}}},\ }\bibfield  {title} {\bibinfo {title} {{DYNESTY: a dynamic
  nested sampling package for estimating Bayesian posteriors and evidences}},\
  }\href {https://doi.org/10.1093/mnras/staa278} {\bibfield  {journal}
  {\bibinfo  {journal} {\mnras}\ }\textbf {\bibinfo {volume} {493}},\ \bibinfo
  {pages} {3132} (\bibinfo {year} {2020})},\ \Eprint
  {https://arxiv.org/abs/1904.02180} {arXiv:1904.02180 [astro-ph.IM]}
  \BibitemShut {NoStop}%
\bibitem [{\citenamefont {Ashton}\ \emph {et~al.}(2019)\citenamefont {Ashton}
  \emph {et~al.}}]{Ashton:2018jfp}%
  \BibitemOpen
  \bibfield  {author} {\bibinfo {author} {\bibfnamefont {G.}~\bibnamefont
  {Ashton}} \emph {et~al.},\ }\bibfield  {title} {\bibinfo {title} {{BILBY: A
  user-friendly Bayesian inference library for gravitational-wave astronomy}},\
  }\href {https://doi.org/10.3847/1538-4365/ab06fc} {\bibfield  {journal}
  {\bibinfo  {journal} {Astrophys. J. Suppl.}\ }\textbf {\bibinfo {volume}
  {241}},\ \bibinfo {pages} {27} (\bibinfo {year} {2019})},\ \Eprint
  {https://arxiv.org/abs/1811.02042} {arXiv:1811.02042 [astro-ph.IM]}
  \BibitemShut {NoStop}%
\bibitem [{\citenamefont {Vallisneri}(2008)}]{Vallisneri:2007ev}%
  \BibitemOpen
  \bibfield  {author} {\bibinfo {author} {\bibfnamefont {M.}~\bibnamefont
  {Vallisneri}},\ }\bibfield  {title} {\bibinfo {title} {{Use and abuse of the
  Fisher information matrix in the assessment of gravitational-wave
  parameter-estimation prospects}},\ }\href
  {https://doi.org/10.1103/PhysRevD.77.042001} {\bibfield  {journal} {\bibinfo
  {journal} {Phys. Rev. D}\ }\textbf {\bibinfo {volume} {77}},\ \bibinfo
  {pages} {042001} (\bibinfo {year} {2008})},\ \Eprint
  {https://arxiv.org/abs/gr-qc/0703086} {arXiv:gr-qc/0703086} \BibitemShut
  {NoStop}%
\bibitem [{\citenamefont {Rodriguez}\ \emph {et~al.}(2014)\citenamefont
  {Rodriguez}, \citenamefont {Farr}, \citenamefont {Raymond}, \citenamefont
  {Farr}, \citenamefont {Littenberg}, \citenamefont {Fazi},\ and\ \citenamefont
  {Kalogera}}]{Rodriguez:2013oaa}%
  \BibitemOpen
  \bibfield  {author} {\bibinfo {author} {\bibfnamefont {C.~L.}\ \bibnamefont
  {Rodriguez}}, \bibinfo {author} {\bibfnamefont {B.}~\bibnamefont {Farr}},
  \bibinfo {author} {\bibfnamefont {V.}~\bibnamefont {Raymond}}, \bibinfo
  {author} {\bibfnamefont {W.~M.}\ \bibnamefont {Farr}}, \bibinfo {author}
  {\bibfnamefont {T.~B.}\ \bibnamefont {Littenberg}}, \bibinfo {author}
  {\bibfnamefont {D.}~\bibnamefont {Fazi}},\ and\ \bibinfo {author}
  {\bibfnamefont {V.}~\bibnamefont {Kalogera}},\ }\bibfield  {title} {\bibinfo
  {title} {{Basic Parameter Estimation of Binary Neutron Star Systems by the
  Advanced LIGO/Virgo Network}},\ }\href
  {https://doi.org/10.1088/0004-637X/784/2/119} {\bibfield  {journal} {\bibinfo
   {journal} {Astrophys. J.}\ }\textbf {\bibinfo {volume} {784}},\ \bibinfo
  {pages} {119} (\bibinfo {year} {2014})},\ \Eprint
  {https://arxiv.org/abs/1309.3273} {arXiv:1309.3273 [astro-ph.HE]}
  \BibitemShut {NoStop}%
\bibitem [{\citenamefont {Pratten}\ \emph {et~al.}(2021)\citenamefont {Pratten}
  \emph {et~al.}}]{Pratten:2020ceb}%
  \BibitemOpen
  \bibfield  {author} {\bibinfo {author} {\bibfnamefont {G.}~\bibnamefont
  {Pratten}} \emph {et~al.},\ }\bibfield  {title} {\bibinfo {title}
  {{Computationally efficient models for the dominant and subdominant harmonic
  modes of precessing binary black holes}},\ }\href
  {https://doi.org/10.1103/PhysRevD.103.104056} {\bibfield  {journal} {\bibinfo
   {journal} {Phys. Rev. D}\ }\textbf {\bibinfo {volume} {103}},\ \bibinfo
  {pages} {104056} (\bibinfo {year} {2021})},\ \Eprint
  {https://arxiv.org/abs/2004.06503} {arXiv:2004.06503 [gr-qc]} \BibitemShut
  {NoStop}%
\bibitem [{\citenamefont {Usman}\ \emph {et~al.}(2019)\citenamefont {Usman},
  \citenamefont {Mills},\ and\ \citenamefont {Fairhurst}}]{Usman:2018imj}%
  \BibitemOpen
  \bibfield  {author} {\bibinfo {author} {\bibfnamefont {S.~A.}\ \bibnamefont
  {Usman}}, \bibinfo {author} {\bibfnamefont {J.~C.}\ \bibnamefont {Mills}},\
  and\ \bibinfo {author} {\bibfnamefont {S.}~\bibnamefont {Fairhurst}},\
  }\bibfield  {title} {\bibinfo {title} {{Constraining the Inclinations of
  Binary Mergers from Gravitational-wave Observations}},\ }\href
  {https://doi.org/10.3847/1538-4357/ab0b3e} {\bibfield  {journal} {\bibinfo
  {journal} {Astrophys. J.}\ }\textbf {\bibinfo {volume} {877}},\ \bibinfo
  {pages} {82} (\bibinfo {year} {2019})},\ \Eprint
  {https://arxiv.org/abs/1809.10727} {arXiv:1809.10727 [gr-qc]} \BibitemShut
  {NoStop}%
\bibitem [{\citenamefont {Chen}\ \emph {et~al.}(2019)\citenamefont {Chen},
  \citenamefont {Vitale},\ and\ \citenamefont {Narayan}}]{Chen:2018omi}%
  \BibitemOpen
  \bibfield  {author} {\bibinfo {author} {\bibfnamefont {H.-Y.}\ \bibnamefont
  {Chen}}, \bibinfo {author} {\bibfnamefont {S.}~\bibnamefont {Vitale}},\ and\
  \bibinfo {author} {\bibfnamefont {R.}~\bibnamefont {Narayan}},\ }\bibfield
  {title} {\bibinfo {title} {{Viewing angle of binary neutron star mergers}},\
  }\href {https://doi.org/10.1103/PhysRevX.9.031028} {\bibfield  {journal}
  {\bibinfo  {journal} {Phys. Rev. X}\ }\textbf {\bibinfo {volume} {9}},\
  \bibinfo {pages} {031028} (\bibinfo {year} {2019})},\ \Eprint
  {https://arxiv.org/abs/1807.05226} {arXiv:1807.05226 [astro-ph.HE]}
  \BibitemShut {NoStop}%
\bibitem [{\citenamefont {Bromm}(2006)}]{Bromm:2005ep}%
  \BibitemOpen
  \bibfield  {author} {\bibinfo {author} {\bibfnamefont {V.}~\bibnamefont
  {Bromm}},\ }\bibfield  {title} {\bibinfo {title} {{High-redshift gamma-ray
  bursts from population III progenitors}},\ }\href
  {https://doi.org/10.1086/500799} {\bibfield  {journal} {\bibinfo  {journal}
  {Astrophys. J.}\ }\textbf {\bibinfo {volume} {642}},\ \bibinfo {pages} {382}
  (\bibinfo {year} {2006})},\ \Eprint {https://arxiv.org/abs/astro-ph/0509303}
  {arXiv:astro-ph/0509303} \BibitemShut {NoStop}%
\bibitem [{\citenamefont {de~Souza}\ \emph {et~al.}(2011)\citenamefont
  {de~Souza}, \citenamefont {Yoshida},\ and\ \citenamefont
  {Ioka}}]{deSouza:2011ea}%
  \BibitemOpen
  \bibfield  {author} {\bibinfo {author} {\bibfnamefont {R.~S.}\ \bibnamefont
  {de~Souza}}, \bibinfo {author} {\bibfnamefont {N.}~\bibnamefont {Yoshida}},\
  and\ \bibinfo {author} {\bibfnamefont {K.}~\bibnamefont {Ioka}},\ }\bibfield
  {title} {\bibinfo {title} {{Population III.1 and III.2 Gamma-Ray Bursts:
  Constraints on the Event Rate for Future Radio and X-ray Surveys}},\ }\href
  {https://doi.org/10.1051/0004-6361/201117242} {\bibfield  {journal} {\bibinfo
   {journal} {Astron. Astrophys.}\ }\textbf {\bibinfo {volume} {533}},\
  \bibinfo {pages} {A32} (\bibinfo {year} {2011})},\ \Eprint
  {https://arxiv.org/abs/1105.2395} {arXiv:1105.2395 [astro-ph.CO]}
  \BibitemShut {NoStop}%
\bibitem [{\citenamefont {Koushiappas}\ and\ \citenamefont
  {Loeb}(2017)}]{Koushiappas:2017kqm}%
  \BibitemOpen
  \bibfield  {author} {\bibinfo {author} {\bibfnamefont {S.~M.}\ \bibnamefont
  {Koushiappas}}\ and\ \bibinfo {author} {\bibfnamefont {A.}~\bibnamefont
  {Loeb}},\ }\bibfield  {title} {\bibinfo {title} {{Maximum redshift of
  gravitational wave merger events}},\ }\href
  {https://doi.org/10.1103/PhysRevLett.119.221104} {\bibfield  {journal}
  {\bibinfo  {journal} {Phys. Rev. Lett.}\ }\textbf {\bibinfo {volume} {119}},\
  \bibinfo {pages} {221104} (\bibinfo {year} {2017})},\ \Eprint
  {https://arxiv.org/abs/1708.07380} {arXiv:1708.07380 [astro-ph.CO]}
  \BibitemShut {NoStop}%
\bibitem [{\citenamefont {Mocz}\ \emph {et~al.}(2020)\citenamefont {Mocz} \emph
  {et~al.}}]{Mocz:2019uyd}%
  \BibitemOpen
  \bibfield  {author} {\bibinfo {author} {\bibfnamefont {P.}~\bibnamefont
  {Mocz}} \emph {et~al.},\ }\bibfield  {title} {\bibinfo {title} {{Galaxy
  formation with BECDM \textendash{} II. Cosmic filaments and first
  galaxies}},\ }\href {https://doi.org/10.1093/mnras/staa738} {\bibfield
  {journal} {\bibinfo  {journal} {Mon. Not. Roy. Astron. Soc.}\ }\textbf
  {\bibinfo {volume} {494}},\ \bibinfo {pages} {2027} (\bibinfo {year}
  {2020})},\ \Eprint {https://arxiv.org/abs/1911.05746} {arXiv:1911.05746
  [astro-ph.CO]} \BibitemShut {NoStop}%
\bibitem [{Note2()}]{Note2}%
  \BibitemOpen
  \bibinfo {note} {Although, we note that there are studies suggesting an
  earlier $(z\gtrsim 50)$~\cite {Trenti:2009cj} or a later $(z\lesssim
  20)$~\cite {Tornatore:2007ds} formation of Pop~III stars.}\BibitemShut
  {Stop}%
\bibitem [{\citenamefont {Inayoshi}\ \emph {et~al.}(2017)\citenamefont
  {Inayoshi}, \citenamefont {Hirai}, \citenamefont {Kinugawa},\ and\
  \citenamefont {Hotokezaka}}]{Inayoshi:2017mrs}%
  \BibitemOpen
  \bibfield  {author} {\bibinfo {author} {\bibfnamefont {K.}~\bibnamefont
  {Inayoshi}}, \bibinfo {author} {\bibfnamefont {R.}~\bibnamefont {Hirai}},
  \bibinfo {author} {\bibfnamefont {T.}~\bibnamefont {Kinugawa}},\ and\
  \bibinfo {author} {\bibfnamefont {K.}~\bibnamefont {Hotokezaka}},\ }\bibfield
   {title} {\bibinfo {title} {{Formation pathway of Population III coalescing
  binary black holes through stable mass transfer}},\ }\href
  {https://doi.org/10.1093/mnras/stx757} {\bibfield  {journal} {\bibinfo
  {journal} {Mon. Not. Roy. Astron. Soc.}\ }\textbf {\bibinfo {volume} {468}},\
  \bibinfo {pages} {5020} (\bibinfo {year} {2017})},\ \Eprint
  {https://arxiv.org/abs/1701.04823} {arXiv:1701.04823 [astro-ph.HE]}
  \BibitemShut {NoStop}%
\bibitem [{\citenamefont {Liu}\ and\ \citenamefont
  {Bromm}(2020{\natexlab{a}})}]{Liu:2020lmi}%
  \BibitemOpen
  \bibfield  {author} {\bibinfo {author} {\bibfnamefont {B.}~\bibnamefont
  {Liu}}\ and\ \bibinfo {author} {\bibfnamefont {V.}~\bibnamefont {Bromm}},\
  }\bibfield  {title} {\bibinfo {title} {{The Population III origin of
  GW190521}},\ }\href {https://doi.org/10.3847/2041-8213/abc552} {\bibfield
  {journal} {\bibinfo  {journal} {Astrophys. J. Lett.}\ }\textbf {\bibinfo
  {volume} {903}},\ \bibinfo {pages} {L40} (\bibinfo {year}
  {2020}{\natexlab{a}})},\ \Eprint {https://arxiv.org/abs/2009.11447}
  {arXiv:2009.11447 [astro-ph.GA]} \BibitemShut {NoStop}%
\bibitem [{\citenamefont {Liu}\ and\ \citenamefont
  {Bromm}(2020{\natexlab{b}})}]{Liu:2020ufc}%
  \BibitemOpen
  \bibfield  {author} {\bibinfo {author} {\bibfnamefont {B.}~\bibnamefont
  {Liu}}\ and\ \bibinfo {author} {\bibfnamefont {V.}~\bibnamefont {Bromm}},\
  }\bibfield  {title} {\bibinfo {title} {{Gravitational waves from Population
  III binary black holes formed by dynamical capture}},\ }\href
  {https://doi.org/10.1093/mnras/staa1362} {\bibfield  {journal} {\bibinfo
  {journal} {Mon. Not. Roy. Astron. Soc.}\ }\textbf {\bibinfo {volume} {495}},\
  \bibinfo {pages} {2475} (\bibinfo {year} {2020}{\natexlab{b}})},\ \Eprint
  {https://arxiv.org/abs/2003.00065} {arXiv:2003.00065 [astro-ph.CO]}
  \BibitemShut {NoStop}%
\bibitem [{\citenamefont {Kinugawa}\ \emph {et~al.}(2020)\citenamefont
  {Kinugawa}, \citenamefont {Nakamura},\ and\ \citenamefont
  {Nakano}}]{Kinugawa:2020ego}%
  \BibitemOpen
  \bibfield  {author} {\bibinfo {author} {\bibfnamefont {T.}~\bibnamefont
  {Kinugawa}}, \bibinfo {author} {\bibfnamefont {T.}~\bibnamefont {Nakamura}},\
  and\ \bibinfo {author} {\bibfnamefont {H.}~\bibnamefont {Nakano}},\
  }\bibfield  {title} {\bibinfo {title} {{Chirp Mass and Spin of Binary Black
  Holes from First Star Remnants}},\ }\href
  {https://doi.org/10.1093/mnras/staa2511} {\bibfield  {journal} {\bibinfo
  {journal} {Mon. Not. Roy. Astron. Soc.}\ }\textbf {\bibinfo {volume} {498}},\
  \bibinfo {pages} {3946} (\bibinfo {year} {2020})},\ \Eprint
  {https://arxiv.org/abs/2005.09795} {arXiv:2005.09795 [astro-ph.HE]}
  \BibitemShut {NoStop}%
\bibitem [{\citenamefont {Tanikawa}\ \emph {et~al.}(2021)\citenamefont
  {Tanikawa}, \citenamefont {Susa}, \citenamefont {Yoshida}, \citenamefont
  {Trani},\ and\ \citenamefont {Kinugawa}}]{Tanikawa:2020cca}%
  \BibitemOpen
  \bibfield  {author} {\bibinfo {author} {\bibfnamefont {A.}~\bibnamefont
  {Tanikawa}}, \bibinfo {author} {\bibfnamefont {H.}~\bibnamefont {Susa}},
  \bibinfo {author} {\bibfnamefont {T.}~\bibnamefont {Yoshida}}, \bibinfo
  {author} {\bibfnamefont {A.~A.}\ \bibnamefont {Trani}},\ and\ \bibinfo
  {author} {\bibfnamefont {T.}~\bibnamefont {Kinugawa}},\ }\bibfield  {title}
  {\bibinfo {title} {{Merger rate density of Population III binary black holes
  below, above, and in the pair-instability mass gap}},\ }\href
  {https://doi.org/10.3847/1538-4357/abe40d} {\bibfield  {journal} {\bibinfo
  {journal} {Astrophys. J.}\ }\textbf {\bibinfo {volume} {910}},\ \bibinfo
  {pages} {30} (\bibinfo {year} {2021})},\ \Eprint
  {https://arxiv.org/abs/2008.01890} {arXiv:2008.01890 [astro-ph.HE]}
  \BibitemShut {NoStop}%
\bibitem [{\citenamefont {Ng}\ \emph {et~al.}(2021)\citenamefont {Ng},
  \citenamefont {Vitale}, \citenamefont {Farr},\ and\ \citenamefont
  {Rodriguez}}]{Ng:2020qpk}%
  \BibitemOpen
  \bibfield  {author} {\bibinfo {author} {\bibfnamefont {K.~K.~Y.}\
  \bibnamefont {Ng}}, \bibinfo {author} {\bibfnamefont {S.}~\bibnamefont
  {Vitale}}, \bibinfo {author} {\bibfnamefont {W.~M.}\ \bibnamefont {Farr}},\
  and\ \bibinfo {author} {\bibfnamefont {C.~L.}\ \bibnamefont {Rodriguez}},\
  }\bibfield  {title} {\bibinfo {title} {{Probing multiple populations of
  compact binaries with third-generation gravitational-wave detectors}},\
  }\href {https://doi.org/10.3847/2041-8213/abf8be} {\bibfield  {journal}
  {\bibinfo  {journal} {Astrophys. J. Lett.}\ }\textbf {\bibinfo {volume}
  {913}},\ \bibinfo {pages} {L5} (\bibinfo {year} {2021})},\ \Eprint
  {https://arxiv.org/abs/2012.09876} {arXiv:2012.09876 [astro-ph.CO]}
  \BibitemShut {NoStop}%
\bibitem [{\citenamefont {Vitale}\ \emph {et~al.}(2017)\citenamefont {Vitale},
  \citenamefont {Gerosa}, \citenamefont {Haster}, \citenamefont
  {Chatziioannou},\ and\ \citenamefont {Zimmerman}}]{Vitale:2017cfs}%
  \BibitemOpen
  \bibfield  {author} {\bibinfo {author} {\bibfnamefont {S.}~\bibnamefont
  {Vitale}}, \bibinfo {author} {\bibfnamefont {D.}~\bibnamefont {Gerosa}},
  \bibinfo {author} {\bibfnamefont {C.-J.}\ \bibnamefont {Haster}}, \bibinfo
  {author} {\bibfnamefont {K.}~\bibnamefont {Chatziioannou}},\ and\ \bibinfo
  {author} {\bibfnamefont {A.}~\bibnamefont {Zimmerman}},\ }\bibfield  {title}
  {\bibinfo {title} {{Impact of Bayesian Priors on the Characterization of
  Binary Black Hole Coalescences}},\ }\href
  {https://doi.org/10.1103/PhysRevLett.119.251103} {\bibfield  {journal}
  {\bibinfo  {journal} {Phys. Rev. Lett.}\ }\textbf {\bibinfo {volume} {119}},\
  \bibinfo {pages} {251103} (\bibinfo {year} {2017})},\ \Eprint
  {https://arxiv.org/abs/1707.04637} {arXiv:1707.04637 [gr-qc]} \BibitemShut
  {NoStop}%
\bibitem [{\citenamefont {Zevin}\ \emph {et~al.}(2020)\citenamefont {Zevin},
  \citenamefont {Berry}, \citenamefont {Coughlin}, \citenamefont
  {Chatziioannou},\ and\ \citenamefont {Vitale}}]{Zevin:2020gxf}%
  \BibitemOpen
  \bibfield  {author} {\bibinfo {author} {\bibfnamefont {M.}~\bibnamefont
  {Zevin}}, \bibinfo {author} {\bibfnamefont {C.~P.~L.}\ \bibnamefont {Berry}},
  \bibinfo {author} {\bibfnamefont {S.}~\bibnamefont {Coughlin}}, \bibinfo
  {author} {\bibfnamefont {K.}~\bibnamefont {Chatziioannou}},\ and\ \bibinfo
  {author} {\bibfnamefont {S.}~\bibnamefont {Vitale}},\ }\bibfield  {title}
  {\bibinfo {title} {{You Can\textquoteright{}t Always Get What You Want: The
  Impact of Prior Assumptions on Interpreting GW190412}},\ }\href
  {https://doi.org/10.3847/2041-8213/aba8ef} {\bibfield  {journal} {\bibinfo
  {journal} {Astrophys. J. Lett.}\ }\textbf {\bibinfo {volume} {899}},\
  \bibinfo {pages} {L17} (\bibinfo {year} {2020})},\ \Eprint
  {https://arxiv.org/abs/2006.11293} {arXiv:2006.11293 [astro-ph.HE]}
  \BibitemShut {NoStop}%
\bibitem [{\citenamefont {Bhagwat}\ \emph {et~al.}(2021)\citenamefont
  {Bhagwat}, \citenamefont {De~Luca}, \citenamefont {Franciolini},
  \citenamefont {Pani},\ and\ \citenamefont {Riotto}}]{Bhagwat:2020bzh}%
  \BibitemOpen
  \bibfield  {author} {\bibinfo {author} {\bibfnamefont {S.}~\bibnamefont
  {Bhagwat}}, \bibinfo {author} {\bibfnamefont {V.}~\bibnamefont {De~Luca}},
  \bibinfo {author} {\bibfnamefont {G.}~\bibnamefont {Franciolini}}, \bibinfo
  {author} {\bibfnamefont {P.}~\bibnamefont {Pani}},\ and\ \bibinfo {author}
  {\bibfnamefont {A.}~\bibnamefont {Riotto}},\ }\bibfield  {title} {\bibinfo
  {title} {{The importance of priors on LIGO-Virgo parameter estimation: the
  case of primordial black holes}},\ }\href
  {https://doi.org/10.1088/1475-7516/2021/01/037} {\bibfield  {journal}
  {\bibinfo  {journal} {JCAP}\ }\textbf {\bibinfo {volume} {01}},\ \bibinfo
  {pages} {037}},\ \Eprint {https://arxiv.org/abs/2008.12320} {arXiv:2008.12320
  [astro-ph.CO]} \BibitemShut {NoStop}%
\bibitem [{\citenamefont {{Vikaeus}}\ \emph {et~al.}(2021)\citenamefont
  {{Vikaeus}}, \citenamefont {{Zackrisson}}, \citenamefont {{Schaerer}},
  \citenamefont {{Visbal}}, \citenamefont {{Fransson}}, \citenamefont
  {{Malhotra}}, \citenamefont {{Rhoads}},\ and\ \citenamefont
  {{Sahl{\'e}n}}}]{Vikaeus2021}%
  \BibitemOpen
  \bibfield  {author} {\bibinfo {author} {\bibfnamefont {A.}~\bibnamefont
  {{Vikaeus}}}, \bibinfo {author} {\bibfnamefont {E.}~\bibnamefont
  {{Zackrisson}}}, \bibinfo {author} {\bibfnamefont {D.}~\bibnamefont
  {{Schaerer}}}, \bibinfo {author} {\bibfnamefont {E.}~\bibnamefont
  {{Visbal}}}, \bibinfo {author} {\bibfnamefont {E.}~\bibnamefont
  {{Fransson}}}, \bibinfo {author} {\bibfnamefont {S.}~\bibnamefont
  {{Malhotra}}}, \bibinfo {author} {\bibfnamefont {J.}~\bibnamefont
  {{Rhoads}}},\ and\ \bibinfo {author} {\bibfnamefont {M.}~\bibnamefont
  {{Sahl{\'e}n}}},\ }\bibfield  {title} {\bibinfo {title} {{Conditions for
  detecting lensed Population III galaxies in blind surveys with the James Webb
  Space Telescope, the Roman Space Telescope and Euclid}},\ }\href@noop {}
  {\bibfield  {journal} {\bibinfo  {journal} {arXiv e-prints}\ ,\ \bibinfo
  {eid} {arXiv:2107.01230}} (\bibinfo {year} {2021})},\ \Eprint
  {https://arxiv.org/abs/2107.01230} {arXiv:2107.01230 [astro-ph.GA]}
  \BibitemShut {NoStop}%
\bibitem [{\citenamefont {{Sun}}\ \emph {et~al.}(2021)\citenamefont {{Sun}},
  \citenamefont {{Mirocha}}, \citenamefont {{Mebane}},\ and\ \citenamefont
  {{Furlanetto}}}]{Sun2021}%
  \BibitemOpen
  \bibfield  {author} {\bibinfo {author} {\bibfnamefont {G.}~\bibnamefont
  {{Sun}}}, \bibinfo {author} {\bibfnamefont {J.}~\bibnamefont {{Mirocha}}},
  \bibinfo {author} {\bibfnamefont {R.~H.}\ \bibnamefont {{Mebane}}},\ and\
  \bibinfo {author} {\bibfnamefont {S.~R.}\ \bibnamefont {{Furlanetto}}},\
  }\bibfield  {title} {\bibinfo {title} {{Revealing the formation histories of
  the first stars with the cosmic near-infrared background}},\ }\href@noop {}
  {\bibfield  {journal} {\bibinfo  {journal} {arXiv e-prints}\ ,\ \bibinfo
  {eid} {arXiv:2107.09324}} (\bibinfo {year} {2021})},\ \Eprint
  {https://arxiv.org/abs/2107.09324} {arXiv:2107.09324 [astro-ph.GA]}
  \BibitemShut {NoStop}%
\bibitem [{\citenamefont {{Tanvir}}\ \emph {et~al.}(2021)\citenamefont
  {{Tanvir}}, \citenamefont {{Le Floc'h}}, \citenamefont {{Christensen}},
  \citenamefont {{Caruana}}, \citenamefont {{Salvaterra}}, \citenamefont
  {{Ghirlanda}}, \citenamefont {{Ciardi}}, \citenamefont {{Maio}},
  \citenamefont {{D'Odorico}}, \citenamefont {{Piedipalumbo}}, \citenamefont
  {{Campana}}, \citenamefont {{Noterdaeme}}, \citenamefont {{Graziani}},
  \citenamefont {{Amati}}, \citenamefont {{Bagoly}}, \citenamefont
  {{Bal{\'a}zs}}, \citenamefont {{Basa}}, \citenamefont {{Behar}},
  \citenamefont {{Bozzo}}, \citenamefont {{De Cia}}, \citenamefont {{Della
  Valle}}, \citenamefont {{De Pasquale}}, \citenamefont {{Frontera}},
  \citenamefont {{Gomboc}}, \citenamefont {{G{\"o}tz}}, \citenamefont
  {{Horvath}}, \citenamefont {{Hudec}}, \citenamefont {{Mereghetti}},
  \citenamefont {{O'Brien}}, \citenamefont {{Osborne}}, \citenamefont
  {{Paltani}}, \citenamefont {{Rosati}}, \citenamefont {{Sergijenko}},
  \citenamefont {{Stanway}}, \citenamefont {{Sz{\'e}csi}}, \citenamefont
  {{Toth}}, \citenamefont {{Urata}}, \citenamefont {{Vergani}},\ and\
  \citenamefont {{Zane}}}]{Tanvir2021}%
  \BibitemOpen
  \bibfield  {author} {\bibinfo {author} {\bibfnamefont {N.~R.}\ \bibnamefont
  {{Tanvir}}}, \bibinfo {author} {\bibfnamefont {E.}~\bibnamefont {{Le
  Floc'h}}}, \bibinfo {author} {\bibfnamefont {L.}~\bibnamefont
  {{Christensen}}}, \bibinfo {author} {\bibfnamefont {J.}~\bibnamefont
  {{Caruana}}}, \bibinfo {author} {\bibfnamefont {R.}~\bibnamefont
  {{Salvaterra}}}, \bibinfo {author} {\bibfnamefont {G.}~\bibnamefont
  {{Ghirlanda}}}, \bibinfo {author} {\bibfnamefont {B.}~\bibnamefont
  {{Ciardi}}}, \bibinfo {author} {\bibfnamefont {U.}~\bibnamefont {{Maio}}},
  \bibinfo {author} {\bibfnamefont {V.}~\bibnamefont {{D'Odorico}}}, \bibinfo
  {author} {\bibfnamefont {E.}~\bibnamefont {{Piedipalumbo}}}, \bibinfo
  {author} {\bibfnamefont {S.}~\bibnamefont {{Campana}}}, \bibinfo {author}
  {\bibfnamefont {P.}~\bibnamefont {{Noterdaeme}}}, \bibinfo {author}
  {\bibfnamefont {L.}~\bibnamefont {{Graziani}}}, \bibinfo {author}
  {\bibfnamefont {L.}~\bibnamefont {{Amati}}}, \bibinfo {author} {\bibfnamefont
  {Z.}~\bibnamefont {{Bagoly}}}, \bibinfo {author} {\bibfnamefont {L.~G.}\
  \bibnamefont {{Bal{\'a}zs}}}, \bibinfo {author} {\bibfnamefont
  {S.}~\bibnamefont {{Basa}}}, \bibinfo {author} {\bibfnamefont
  {E.}~\bibnamefont {{Behar}}}, \bibinfo {author} {\bibfnamefont
  {E.}~\bibnamefont {{Bozzo}}}, \bibinfo {author} {\bibfnamefont
  {A.}~\bibnamefont {{De Cia}}}, \bibinfo {author} {\bibfnamefont
  {M.}~\bibnamefont {{Della Valle}}}, \bibinfo {author} {\bibfnamefont
  {M.}~\bibnamefont {{De Pasquale}}}, \bibinfo {author} {\bibfnamefont
  {F.}~\bibnamefont {{Frontera}}}, \bibinfo {author} {\bibfnamefont
  {A.}~\bibnamefont {{Gomboc}}}, \bibinfo {author} {\bibfnamefont
  {D.}~\bibnamefont {{G{\"o}tz}}}, \bibinfo {author} {\bibfnamefont
  {I.}~\bibnamefont {{Horvath}}}, \bibinfo {author} {\bibfnamefont
  {R.}~\bibnamefont {{Hudec}}}, \bibinfo {author} {\bibfnamefont
  {S.}~\bibnamefont {{Mereghetti}}}, \bibinfo {author} {\bibfnamefont {P.~T.}\
  \bibnamefont {{O'Brien}}}, \bibinfo {author} {\bibfnamefont {J.~P.}\
  \bibnamefont {{Osborne}}}, \bibinfo {author} {\bibfnamefont {S.}~\bibnamefont
  {{Paltani}}}, \bibinfo {author} {\bibfnamefont {P.}~\bibnamefont {{Rosati}}},
  \bibinfo {author} {\bibfnamefont {O.}~\bibnamefont {{Sergijenko}}}, \bibinfo
  {author} {\bibfnamefont {E.~R.}\ \bibnamefont {{Stanway}}}, \bibinfo {author}
  {\bibfnamefont {D.}~\bibnamefont {{Sz{\'e}csi}}}, \bibinfo {author}
  {\bibfnamefont {L.~V.}\ \bibnamefont {{Toth}}}, \bibinfo {author}
  {\bibfnamefont {Y.}~\bibnamefont {{Urata}}}, \bibinfo {author} {\bibfnamefont
  {S.}~\bibnamefont {{Vergani}}},\ and\ \bibinfo {author} {\bibfnamefont
  {S.}~\bibnamefont {{Zane}}},\ }\bibfield  {title} {\bibinfo {title}
  {{Exploration of the high-redshift universe enabled by THESEUS}},\
  }\href@noop {} {\bibfield  {journal} {\bibinfo  {journal} {arXiv e-prints}\
  ,\ \bibinfo {eid} {arXiv:2104.09532}} (\bibinfo {year} {2021})},\ \Eprint
  {https://arxiv.org/abs/2104.09532} {arXiv:2104.09532 [astro-ph.IM]}
  \BibitemShut {NoStop}%
\bibitem [{\citenamefont {Mukherjee}\ \emph {et~al.}(2021)\citenamefont
  {Mukherjee}, \citenamefont {Meinema},\ and\ \citenamefont
  {Silk}}]{Mukherjee:2021itf}%
  \BibitemOpen
  \bibfield  {author} {\bibinfo {author} {\bibfnamefont {S.}~\bibnamefont
  {Mukherjee}}, \bibinfo {author} {\bibfnamefont {M.~S.~P.}\ \bibnamefont
  {Meinema}},\ and\ \bibinfo {author} {\bibfnamefont {J.}~\bibnamefont
  {Silk}},\ }\bibfield  {title} {\bibinfo {title} {{Prospects of discovering
  sub-solar primordial black holes using the stochastic gravitational wave
  background from third-generation detectors}},\ }\href@noop {} {\bibfield
  {journal} {\bibinfo  {journal} {arXiv}\ } (\bibinfo {year} {2021})},\ \Eprint
  {https://arxiv.org/abs/2107.02181} {arXiv:2107.02181 [astro-ph.CO]}
  \BibitemShut {NoStop}%
\bibitem [{\citenamefont {Farr}\ \emph {et~al.}(2015)\citenamefont {Farr},
  \citenamefont {Gair}, \citenamefont {Mandel},\ and\ \citenamefont
  {Cutler}}]{Farr:2013yna}%
  \BibitemOpen
  \bibfield  {author} {\bibinfo {author} {\bibfnamefont {W.~M.}\ \bibnamefont
  {Farr}}, \bibinfo {author} {\bibfnamefont {J.~R.}\ \bibnamefont {Gair}},
  \bibinfo {author} {\bibfnamefont {I.}~\bibnamefont {Mandel}},\ and\ \bibinfo
  {author} {\bibfnamefont {C.}~\bibnamefont {Cutler}},\ }\bibfield  {title}
  {\bibinfo {title} {{Counting And Confusion: Bayesian Rate Estimation With
  Multiple Populations}},\ }\href {https://doi.org/10.1103/PhysRevD.91.023005}
  {\bibfield  {journal} {\bibinfo  {journal} {Phys. Rev. D}\ }\textbf {\bibinfo
  {volume} {91}},\ \bibinfo {pages} {023005} (\bibinfo {year} {2015})},\
  \Eprint {https://arxiv.org/abs/1302.5341} {arXiv:1302.5341 [astro-ph.IM]}
  \BibitemShut {NoStop}%
\bibitem [{\citenamefont {Wysocki}\ \emph {et~al.}(2019)\citenamefont
  {Wysocki}, \citenamefont {Lange},\ and\ \citenamefont
  {O'Shaughnessy}}]{Wysocki:2018mpo}%
  \BibitemOpen
  \bibfield  {author} {\bibinfo {author} {\bibfnamefont {D.}~\bibnamefont
  {Wysocki}}, \bibinfo {author} {\bibfnamefont {J.}~\bibnamefont {Lange}},\
  and\ \bibinfo {author} {\bibfnamefont {R.}~\bibnamefont {O'Shaughnessy}},\
  }\bibfield  {title} {\bibinfo {title} {{Reconstructing phenomenological
  distributions of compact binaries via gravitational wave observations}},\
  }\href {https://doi.org/10.1103/PhysRevD.100.043012} {\bibfield  {journal}
  {\bibinfo  {journal} {Phys. Rev. D}\ }\textbf {\bibinfo {volume} {100}},\
  \bibinfo {pages} {043012} (\bibinfo {year} {2019})},\ \Eprint
  {https://arxiv.org/abs/1805.06442} {arXiv:1805.06442 [gr-qc]} \BibitemShut
  {NoStop}%
\bibitem [{\citenamefont {Thrane}\ and\ \citenamefont
  {Talbot}(2019)}]{Thrane:2018qnx}%
  \BibitemOpen
  \bibfield  {author} {\bibinfo {author} {\bibfnamefont {E.}~\bibnamefont
  {Thrane}}\ and\ \bibinfo {author} {\bibfnamefont {C.}~\bibnamefont
  {Talbot}},\ }\bibfield  {title} {\bibinfo {title} {{An introduction to
  Bayesian inference in gravitational-wave astronomy: parameter estimation,
  model selection, and hierarchical models}},\ }\href
  {https://doi.org/10.1017/pasa.2019.2} {\bibfield  {journal} {\bibinfo
  {journal} {Publ. Astron. Soc. Austral.}\ }\textbf {\bibinfo {volume} {36}},\
  \bibinfo {pages} {e010} (\bibinfo {year} {2019})},\ \bibinfo {note}
  {[Erratum: Publ.Astron.Soc.Austral. 37, e036 (2020)]},\ \Eprint
  {https://arxiv.org/abs/1809.02293} {arXiv:1809.02293 [astro-ph.IM]}
  \BibitemShut {NoStop}%
\bibitem [{\citenamefont {Mandel}\ \emph {et~al.}(2019)\citenamefont {Mandel},
  \citenamefont {Farr},\ and\ \citenamefont {Gair}}]{Mandel:2018mve}%
  \BibitemOpen
  \bibfield  {author} {\bibinfo {author} {\bibfnamefont {I.}~\bibnamefont
  {Mandel}}, \bibinfo {author} {\bibfnamefont {W.~M.}\ \bibnamefont {Farr}},\
  and\ \bibinfo {author} {\bibfnamefont {J.~R.}\ \bibnamefont {Gair}},\
  }\bibfield  {title} {\bibinfo {title} {{Extracting distribution parameters
  from multiple uncertain observations with selection biases}},\ }\href
  {https://doi.org/10.1093/mnras/stz896} {\bibfield  {journal} {\bibinfo
  {journal} {Mon. Not. Roy. Astron. Soc.}\ }\textbf {\bibinfo {volume} {486}},\
  \bibinfo {pages} {1086} (\bibinfo {year} {2019})},\ \Eprint
  {https://arxiv.org/abs/1809.02063} {arXiv:1809.02063 [physics.data-an]}
  \BibitemShut {NoStop}%
\bibitem [{\citenamefont {Vitale}(2020)}]{Vitale:2020aaz}%
  \BibitemOpen
  \bibfield  {author} {\bibinfo {author} {\bibfnamefont {S.}~\bibnamefont
  {Vitale}},\ }\bibfield  {title} {\bibinfo {title} {{One, No One, and One
  Hundred Thousand -- Inferring the properties of a population in presence of
  selection effects}},\ }\href@noop {} {\bibfield  {journal} {\bibinfo
  {journal} {arXiv e-print}\ } (\bibinfo {year} {2020})},\ \Eprint
  {https://arxiv.org/abs/2007.05579} {arXiv:2007.05579 [astro-ph.IM]}
  \BibitemShut {NoStop}%
\bibitem [{\citenamefont {Aghanim}\ \emph {et~al.}(2020)\citenamefont {Aghanim}
  \emph {et~al.}}]{Planck:2018vyg}%
  \BibitemOpen
  \bibfield  {author} {\bibinfo {author} {\bibfnamefont {N.}~\bibnamefont
  {Aghanim}} \emph {et~al.} (\bibinfo {collaboration} {Planck}),\ }\bibfield
  {title} {\bibinfo {title} {{Planck 2018 results. VI. Cosmological
  parameters}},\ }\href {https://doi.org/10.1051/0004-6361/201833910}
  {\bibfield  {journal} {\bibinfo  {journal} {Astron. Astrophys.}\ }\textbf
  {\bibinfo {volume} {641}},\ \bibinfo {pages} {A6} (\bibinfo {year} {2020})},\
  \Eprint {https://arxiv.org/abs/1807.06209} {arXiv:1807.06209 [astro-ph.CO]}
  \BibitemShut {NoStop}%
\bibitem [{\citenamefont {Trenti}\ and\ \citenamefont
  {Stiavelli}(2009)}]{Trenti:2009cj}%
  \BibitemOpen
  \bibfield  {author} {\bibinfo {author} {\bibfnamefont {M.}~\bibnamefont
  {Trenti}}\ and\ \bibinfo {author} {\bibfnamefont {M.}~\bibnamefont
  {Stiavelli}},\ }\bibfield  {title} {\bibinfo {title} {{The Formation Rates of
  Population III Stars and Chemical Enrichment of Halos during the Reionization
  Era}},\ }\href {https://doi.org/10.1088/0004-637X/694/2/879} {\bibfield
  {journal} {\bibinfo  {journal} {Astrophys. J.}\ }\textbf {\bibinfo {volume}
  {694}},\ \bibinfo {pages} {879} (\bibinfo {year} {2009})},\ \Eprint
  {https://arxiv.org/abs/0901.0711} {arXiv:0901.0711 [astro-ph.CO]}
  \BibitemShut {NoStop}%
\bibitem [{\citenamefont {Tornatore}\ \emph {et~al.}(2007)\citenamefont
  {Tornatore}, \citenamefont {Borgani}, \citenamefont {Dolag},\ and\
  \citenamefont {Matteucci}}]{Tornatore:2007ds}%
  \BibitemOpen
  \bibfield  {author} {\bibinfo {author} {\bibfnamefont {L.}~\bibnamefont
  {Tornatore}}, \bibinfo {author} {\bibfnamefont {S.}~\bibnamefont {Borgani}},
  \bibinfo {author} {\bibfnamefont {K.}~\bibnamefont {Dolag}},\ and\ \bibinfo
  {author} {\bibfnamefont {F.}~\bibnamefont {Matteucci}},\ }\bibfield  {title}
  {\bibinfo {title} {{Chemical enrichment of galaxy clusters from
  hydrodynamical simulations}},\ }\href
  {https://doi.org/10.1111/j.1365-2966.2007.12070.x} {\bibfield  {journal}
  {\bibinfo  {journal} {Mon. Not. Roy. Astron. Soc.}\ }\textbf {\bibinfo
  {volume} {382}},\ \bibinfo {pages} {1050} (\bibinfo {year} {2007})},\ \Eprint
  {https://arxiv.org/abs/0705.1921} {arXiv:0705.1921 [astro-ph]} \BibitemShut
  {NoStop}%
\end{thebibliography}%
